\documentclass{svmult}

\usepackage{type1cm}        
\usepackage{nicefrac}
\usepackage{makeidx}         
\usepackage{graphicx}        
\usepackage{multicol}        
\usepackage[bottom]{footmisc}

\usepackage{newtxtext}       %
\usepackage[varvw]{newtxmath}       

\usepackage{cprotect}

\usepackage{color}
\usepackage{natbib}
\usepackage{xspace}
\usepackage{sidecap}
\usepackage{ifthen}
\usepackage[colorlinks = true, urlcolor = blue, citecolor = blue]{hyperref}

\newcommand{\msun}{M_{\hbox{$\odot$}}\xspace}
\newcommand{\zsun}{$Z_{\hbox{$\odot$}}$\xspace}
\newcommand{\sun}{{\hbox{$\odot$}}\xspace}
\newcommand{\kms}{\hbox{km s$^{-1}$}\xspace}
\newcommand{\lx}{L_{\rm X}}

\newcommand{\chandra}{\emph{Chandra}\xspace}
\newcommand{\xmm}{\emph{XMM--Newton}\xspace}
\newcommand{\einstein}{\emph{Einstein}\xspace}
\newcommand{\athena}{\emph{Athena}\xspace}
\newcommand{\lynx}{\emph{Lynx}\xspace}

\newcommand*{\Mdotbh}{{\dot M}_{\rm BH}}
\newcommand*{\Lbh}{L_{\rm rad}}
\newcommand*{\Lw}{L_{\rm w}}
\newcommand*{\epsw}{\epsilon_{\rm w}}
\newcommand*{\epsem}{\epsilon_{\rm rad}}

\makeindex             

\def\thechapter{\vspace*{-2pc}}
\def\chaptername{}
\begin{document}

\title*{The Hot Interstellar Medium}

\author{Emanuele Nardini, Dong-Woo Kim, and Silvia Pellegrini}

\institute{Emanuele Nardini \at INAF\,--\,Arcetri Astrophysical Observatory, Largo Enrico Fermi 5, I-50125 Firenze, Italy, \email{emanuele.nardini@inaf.it}
\and
Dong-Woo Kim \at Center for Astrophysics | Harvard \& Smithsonian, 60 Garden Street, Cambridge, MA 02138, USA,  \email{dkim@cfa.harvard.edu}
\and
Silvia Pellegrini \at Department of Physics and Astronomy, University of Bologna, via Piero Gobetti 93/2, I-40129 Bologna, Italy, \email{silvia.pellegrini@unibo.it}}

\titlerunning{The Hot ISM}
\authorrunning{Nardini, Kim, Pellegrini}

\maketitle

\begin{refguide}

\begin{sloppy}

\def\thechapter{\arabic{chapter}}


\abstract{The interstellar medium (ISM) of galaxies very often contains a gas component that reaches the temperature of several million degrees, whose physical and chemical properties can be investigated through imaging and spectroscopy in the X-rays. We review the current knowledge on the origin and retention of the hot ISM in star-forming and early-type galaxies, from a combined theoretical and observational standpoint. As a complex interplay between gravitational processes, environmental effects, and feedback mechanisms contributes to its physical conditions, the hot ISM represents a key diagnostic of the evolution of galaxies.}

{\bf Keywords}~X-rays: ISM; X-rays: galaxies; ISM: evolution; galaxies: ISM; galaxies: halos; galaxies: starburst; galaxies: elliptical and lenticular, cD; galaxies: evolution

\section{Introduction}
\label{intro}
One of the most surprising discoveries that came with the first images of galaxies in the X-ray band was the ubiquitous presence of diffuse emission, suggesting that a substantial fraction of the interstellar medium (ISM) is heated to temperatures of $\sim$10$^6$--10$^7$ K, regardless of the galaxy type. A compelling question immediately emerged: what is the physical origin of this hot gas? At first, the solution did not appear to be straightforward, as no obvious source of energy input into the ISM (from, e.g., fierce star formation, or an active galactic nucleus, AGN) was systematically involved, especially in largely passive objects like early-type galaxies. We now know that the processes responsible for the creation and subsequent maintenance of the hot ISM are manifold, and that their relative importance widely varies with the different galaxy classes. 

In star-forming galaxies, the luminosity of the hot ISM is proportional to the rate at which stars are assembled. The energy released by young, massive stars, while they age and when they finally explode as supernovae (SNe), can easily prevail over gravity and drive the expansion of a hot, high-pressure bubble into the cold ISM, which is shock-heated and swept-up. The resulting large-scale wind also functions as a carrier of the heavy elements synthesized in the stellar nurseries, and as a feedback force whose activity can explain several properties of the global galaxy population. 

In early-type systems, instead, the hot ISM evolves under the effect of the gravitational field of the host galaxy and dark matter, of various forms of mass exchange with the galaxy and its environment (as provided by stellar evolution input, gas accretion from the circumgalactic medium, and mass deposition via cooling), of major energy sources and radiative losses. Heating is provided by SN explosions, thermalization of the random and ordered kinetic energy of the stellar motions, and accretion onto the central supermassive black hole (SMBH) followed by a number of possible forms of energy injection. 

As star-forming and early-type galaxies can be thought of as the opposite ends of the same evolutionary sequence, linked through mergers in a hierarchical scenario, the overview we provide here practically covers the full life cycle of the hot ISM. 

This Chapter is organized as follows. 

Section \ref{sfg} is dedicated to star-forming galaxies, and illustrates the connection between the hot ISM and the mass and energy deposited by core-collapse SNe. We describe the consequences of feedback from starburst-driven winds and their crucial role in the chemical enrichment of galaxy halos and intergalactic medium. 

Section \ref{etg-k} deals with the main properties of the hot ISM that can be derived from the X-ray observations of early-type galaxies. We discuss how the scaling relations between the various physical quantities and their spatial distributions compare with the predictions of theoretical models. 

Section \ref{etg-p} examines the origin of the hot gas in early-type galaxies, its heating sources, dynamical status, and relationship with the galactic properties. We outline a global picture of the hot gas origin and evolution in these galaxies, which takes into account the constraints from X-ray observations. 

Section \ref{fut} briefly summarizes the future prospects of this research field.

\section{The Hot ISM of Star-Forming Galaxies}
\label{sfg}
The birth, evolution, and death of stars constitute a key process in the lifetime of a galaxy, which eventually returns energy---in the form of radiation and winds---and metals to the ISM. The basic requirements for star formation are the availability of an ample reservoir of cold, dense gas, and the existence of internal and/or external agents to prompt its gravitational collapse. Both conditions are usually met in late-type (spiral), dwarf/irregular, and interacting galaxies. The impact of star formation on the host environment is mostly dramatic during the so-called `starburst' phases. In a starburst galaxy (after \citealt{Weedman+81}), star formation is currently much more intense than in the past, as revealed by the fact that the ratio between stellar mass ($M_\star$) and star formation rate (SFR) is significantly smaller than the Hubble time.\footnote{The most commonly used quantity is actually the reciprocal ratio SFR/$M_\star$, which is named `specific star formation rate' (sSFR).} In turn, this implies that the ongoing star formation episode must be short-lived (hence the notion of a burst), as it cannot be sustained over a long period without consuming the gas supply of the whole galaxy.

Our own Galaxy, the Milky Way, has a SFR of $\approx$2 $\msun$ yr$^{-1}$ (see \citealt{Chomiuk-Povich_11}, and references therein), and the total atomic plus molecular gas mass in the disk plane within a Galactocentric radius of 13.5 kpc is $M_{\rm GAS} \simeq 6.6 \times 10^9~\msun$ \citep{Kennicutt-Evans_12}. Consequently, the gas depletion timescale ($t_{\rm dep} = M_{\rm GAS}/{\rm SFR}$) is of the order of a few Gyr, qualifying the Milky Way as a `normal' star-forming galaxy. By contrast, the star formation efficiency ($= 1/t_{\rm dep}$) of our largest companion in the Local Group, the Andromeda Galaxy (M\,31), is remarkably lower (${\rm SFR} \sim 0.3~\msun$ yr$^{-1}$; \citealt{Tabatabaei-Berkhuijsen_10}), more similar to the quiescent, early-type galaxies (ETGs) that are discussed in the second part of this Chapter. The presence of X-ray bubbles in the halo of the Milky Way \citep{Predehl+20} proves beyond doubt that even in normal galaxies the hot ISM keeps a record of any past activity. Hence, without loss of generality, in this Section we will focus on galaxies that are now actively star-forming, i.e., those in which star formation has been significantly enhanced by some recent form of instability or perturbation. These objects typically emit most of their luminosity in the infrared, as the light of newly formed stars is almost entirely reprocessed by dust. 

\subsection{Shock-Heating and Diffuse X-ray Emission}
\label{sfg-1}
The terminal stages of stellar evolution are characterized by violent, high-energy events that naturally lead to copious X-ray emission. Several different mechanisms can contribute to the X-ray spectrum of a starburst galaxy (see \citealt{Persic-Rephaeli_02} for a concise review), including accretion onto a compact object in high- and low-mass X-ray binaries, shock-heating and synchrotron radiation from young supernova remnants, boosted Compton up-scattering of infrared photons by ambient relativistic electrons. Diffuse, thermal X-ray emission is also expected due to the combined shock-heating capacity of stellar winds (from, e.g., Wolf-Rayet stars) and SN explosions. In the limit of strong, adiabatic shocks, the relation between shock velocity ($v_{\rm sh}$) and post-shock temperature ($T_{\rm sh}$) is given by: 
\begin{equation}
\label{eq.01}
kT_{\rm sh} \simeq 3 \mu m_p v_{\rm sh}^2 /16,
\end{equation}
where $k$ is Boltzmann's constant, $\mu$ the mean molecular weight, and $m_p$ the proton mass. For $\mu \simeq 0.6$, as per fully ionized gas with solar abundances, and $v_{\rm sh} \approx$ 1,000 \kms (Section \ref{sfg-2}), the ISM can be heated to temperatures of the order of $\sim$1 keV.

In general, all of the above components have comparable weights, making it difficult to spectrally disentangle them when spatially mixed, in spite of their different shapes. The advent of sub-arcsecond resolution with the \chandra X-ray observatory \citep{Weisskopf+02} eventually allowed astronomers to identify all the point sources in nearby galaxies down to luminosities of $\lx \approx 10^{37}$ erg s$^{-1}$ (see, for instance, the census of discrete X-ray sources in NGC\,4278 by \citealt{Brassington+09}). The possibility of removing this contamination to the diffuse X-ray emission represented a huge leap forward in the study of the physical properties of the hot ISM. 

The existence of extended X-ray emission with complex morphological features in star-forming galaxies had been known since the \einstein mission \citep{Giacconi+79}, the first X-ray telescope with full imaging capabilities. In particular, the observation of M\,82 (the Cigar Galaxy, at $D \sim 3.6$ Mpc of distance) unveiled a patchy halo extending for several kpc along the minor axis of the highly inclined galactic plane, above and below the nuclear region \citep{Watson+84}. Intense, diffuse emission was also detected in the disk, closely matching the optical nucleus. The peculiar morphology of the X-ray halo, whose base coincides with the nuclear starburst, and the tight spatial correlation with the H$\alpha$ filaments in the optical halo, supported the notion that the hot gas was in an outflow state. Similar conclusions were drawn for NGC\,253 (the Sculptor Galaxy, $D \sim 3.5$ Mpc), whose \einstein image emphasized a plume-like structure protruding southward from the nucleus along the minor axis of the disk \citep{Fabbiano-Trinchieri_84}.

\begin{figure*}[b!]
\centering
\includegraphics[width=0.85\textwidth]{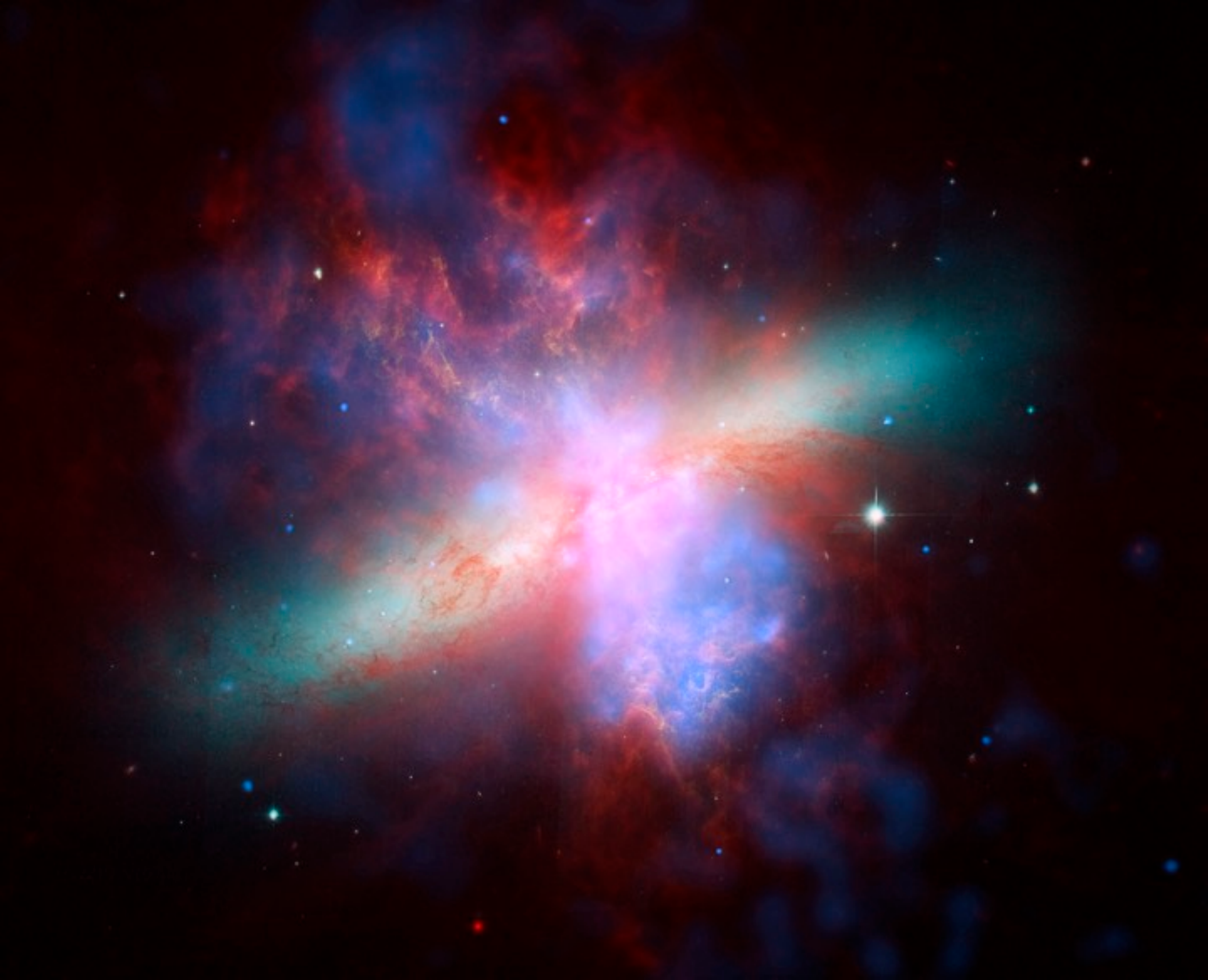}
\caption{Multi-wavelength image of the starburst galaxy M\,82, obtained by combining the data from \chandra (blue: hot gas), {\it Hubble} (green: starlight, orange: H$\alpha$), and {\it Spitzer} (red: cold gas and dust). While the optical semblance is that of a normal spiral galaxy, the broadband view immediately conveys the impression of a tremendous explosion taking place in the nuclear region. The ensuing wind is a complex, multi-phase medium, as evinced by the co-existence between million-K gas and cold dust grains. Credits: NASA/CXC/JHU/D.\,Strickland (X-ray), NASA/ESA/STScI/AURA/The Hubble Heritage Team (optical), NASA/JPL--Caltech/Univ.\,of\,AZ/C.\,Engelbracht (infrared).}
\label{fig.01}
\end{figure*}

The remarkable overlap between extra-planar soft X-ray and optical (H$\alpha$) emission is now known to be a distinctive attribute of edge-on, star-forming galaxies in the local Universe. This turned out to be instrumental in establishing the connection of the diffuse X-ray emission to the starburst wind phenomenon. In fact, while the X-ray study of the hot ISM was still in a pioneering phase, extensive imaging and spectroscopic observations in the optical had already provided a large body of evidence in favour of the wind model \citep{Lehnert-Heckman_96}, taking advantage of the higher sensitivity and much finer spatial and spectral resolution, which allowed a superior grasp on the morphological, dynamical, and physical properties of the gas. The merits of a multi-wavelength perspective for achieving a comprehensive picture of the physics involved are illustrated in Figure \ref{fig.01}, which displays the composite image of M\,82 as obtained by the {\it Hubble} (optical), {\it Spitzer} (infrared), and \chandra (X-ray) space observatories. In this framework, the spatial coincidence between soft X-ray and H$\alpha$ emission does not follow from the cooling of the hot, fast wind, but from its interaction with the inhomogeneous ISM in the galaxy halo \citep[e.g.,][]{Lehnert+99,Strickland+02}. When the wind impinges on a cold, dense cloud, the slow, forward shock propagating into the cloud can heat the gas to temperatures $T \sim 10^4$ K, leading to H$\alpha$ emission via recombination, whereas the fast, reverse shock into the wind would simultaneously produce the X-rays (Equation \ref{eq.01}). This is better understood by revisiting the basic physics of starburst winds.

\subsection{Theory and Observations of Superwinds}
\label{sfg-2}
Massive stars and SNe can power galaxy-wide outflows, informally known as `superwinds', when the kinetic energy of their ejecta is efficiently thermalized. The simplest analytical treatment of a superwind assumes spherical symmetry and neglects gravity, radiative cooling, and ambient gas \citep{Chevalier-Clegg_85}. The starburst is described as a central region of radius $R_{\rm SB}$ with constant mass ($\dot M_{\rm SB}$) and energy ($\dot E_{\rm SB}$) injection rates. Before discussing the solutions of the hydrodynamical equations for this `free-flowing' wind, it is useful to derive order-of-magnitude estimates for both $\dot M_{\rm SB}$ and $\dot E_{\rm SB}$. Considering only core-collapse SNe, which are the most relevant to a starburst phase (Section \ref{sfg-3}), the standard values of the mass and energy deposited by a single SN are 10 $\msun$ and 10$^{51}$ erg, respectively, while the scaling between SN rate and SFR is $\sim$0.01 $\msun^{-1}$. Hence, the characteristic mass and energy injection rates are:
\begin{equation}
\dot M_{\rm SB} \approx 6.3 \times 10^{24}\, ({\rm SFR}/\msun\,{\rm yr}^{-1})\quad {\rm g\ s}^{-1},
\label{eq.02}
\end{equation}
and
\begin{equation}
\label{eq.03}
\dot E_{\rm SB} \approx 3.2 \times 10^{41}\, ({\rm SFR}/\msun\,{\rm yr}^{-1})\quad {\rm erg\ s}^{-1}.
\end{equation}
These numerical expressions will be used in the remainder of this Section.

In the free-wind model, the kinetic energy of the colliding SN ejecta and stellar winds is thermalized via shocks. Radiative losses are indeed negligible, given the high temperatures and low densities involved. The complete solution to the flow equations\footnote{A detailed analytical derivation of the solution, applied to dense clusters of massive stars, can be found in \citet{Canto+00}.} implies a central ($r \ll R_{\rm SB}$) temperature of the order of $\sim$10$^8$ K:
\begin{equation}
\label{eq.04}
kT_{\rm SB} = 0.4 \mu m_p \, (\dot E_{\rm SB}/\dot M_{\rm SB}).
\end{equation}
The central density of this hot gas is given by:
\begin{equation}
\label{eq.05}
\rho_{\rm SB} \simeq 0.3 \, \dot M_{\rm SB}^{3/2} \, \dot E_{\rm SB}^{-1/2} \, R_{\rm SB}^{-2},
\end{equation}
which, for $R_{\rm SB} = 200$ pc (as originally tailored to M\,82), corresponds to a number density $n_{\rm SB} \sim 0.02\,({\rm SFR}/\msun\,{\rm yr}^{-1})$ cm$^{-3}$. The high-pressure ($P_{\rm SB} \sim 2n_{\rm SB}kT_{\rm SB} > 10^{-9}$ dyn cm$^{-2}$), hot bubble expands adiabatically into the surrounding ISM, becoming supersonic at $r = R_{\rm SB}$. At larger distance ($r \gg R_{\rm SB}$), the wind temperature and density respectively decline as $\propto r^{-4/3}$ and $\propto r^{-2}$. The velocity rapidly converges to the terminal value, which can be recovered by imposing the equality of energy injection rate and asymptotic kinetic energy flux: 
\begin{equation}
\label{eq.06}	
v_{\rm wind} = (2 \dot E_{\rm SB} / \dot M_{\rm SB})^{1/2},
\end{equation}
i.e., $\approx$3,000 \kms. As the typical escape velocity ($v_{\rm esc}$) from spirals is around 500 \kms, neglecting gravity is then confirmed to be a fair working assumption. 

In a more realistic case (see \citealt{Veilleux+05,Zhang_18}), one must include a thermalization efficiency ($\xi$) and a mass-loading factor ($\beta$): the former accounts for radiative losses within the starburst region, while the latter allows for the amplification of the intrinsic mass injection rate due to the evaporation of cold ISM clouds \citep{Suchkov+96}. Intuitively, if $\xi < 1$ and/or $\beta > 1$, the wind is slowed down. The terminal velocity of the wind in Equation \ref{eq.06} should therefore be corrected by a $(\xi/\beta)^{1/2}$ term. The thermalization efficiency is always expected to be rather high. Different hydrodynamical models of the superwind in M\,82 comply with the observational constraints for $\xi \geq 0.3$ \citep{Strickland-Heckman_09}. This is empirically corroborated by the relation between intrinsic bolometric luminosity of the hot ISM and SFR obtained for a sample of 21 star-forming galaxies by \citet{Mineo+12}, $L_{\rm bol}/{\rm SFR} \sim 1.5 \times 10^{40} ~ {\rm erg\ s}^{-1} ~ (\msun\ {\rm yr}^{-1})^{-1}$, weighing the degree of radiative losses (cf. Equation \ref{eq.03}). Mass loading can instead be considerable (up to $\beta \sim 10$; \citealt{Zhang+14}), so that the velocity of real winds rarely exceeds $\sim$1,000 \kms.

Although the surface brightness profile of the X-ray halo in M\,82 was suggested to be consistent with a free wind \citep{Fabbiano_88}, the presumed emissivity (which goes like $\propto n^2$) is thought to be too low for the observed X-rays to be directly emitted from the hot wind itself. Measuring the density of the hot, X-ray emitting gas is not a trivial task. In the most popular spectral models of X-ray plasmas the density is encapsulated in a quantity called `emission measure', defined as ${\rm EM} = \int n_e n_{\rm H} dV \sim n_e^2 \eta V$, where $n_e$ and $n_{\rm H}$ are the number densities of electrons and hydrogen ions, $V$ is the geometrical volume occupied by the gas, and $\eta$ is the (unknown) filling factor. Numerical models suggest that the hot ($T \sim 10^7$ K) phase, which can be identified with the free wind, is volume-filling, but it barely contributes to the observed X-ray emission because of its low density ($n_e \sim 0.01$ cm$^{-3}$; \citealt{Strickland-Stevens_00}). Only within the starburst region, where densities are higher (Equation \ref{eq.05}), such a component becomes detectable \citep{Strickland-Heckman_07}. Most of the soft X-ray emission from superwinds is predicted to originate from gas with higher density and much lower filling factor, in agreement with the `shocked-cloud' scenario. The uncertainty on $\eta$, however, heavily affects several key properties of the hot ISM that can be extracted from the X-ray spectra (e.g., gas mass, pressure, thermal energy). 

\subsection{Chemical and Physical Evolution of the Hot ISM}
\label{sfg-3}
%
\begin{figure*}[t!]
\centering
\includegraphics[width=0.85\textwidth]{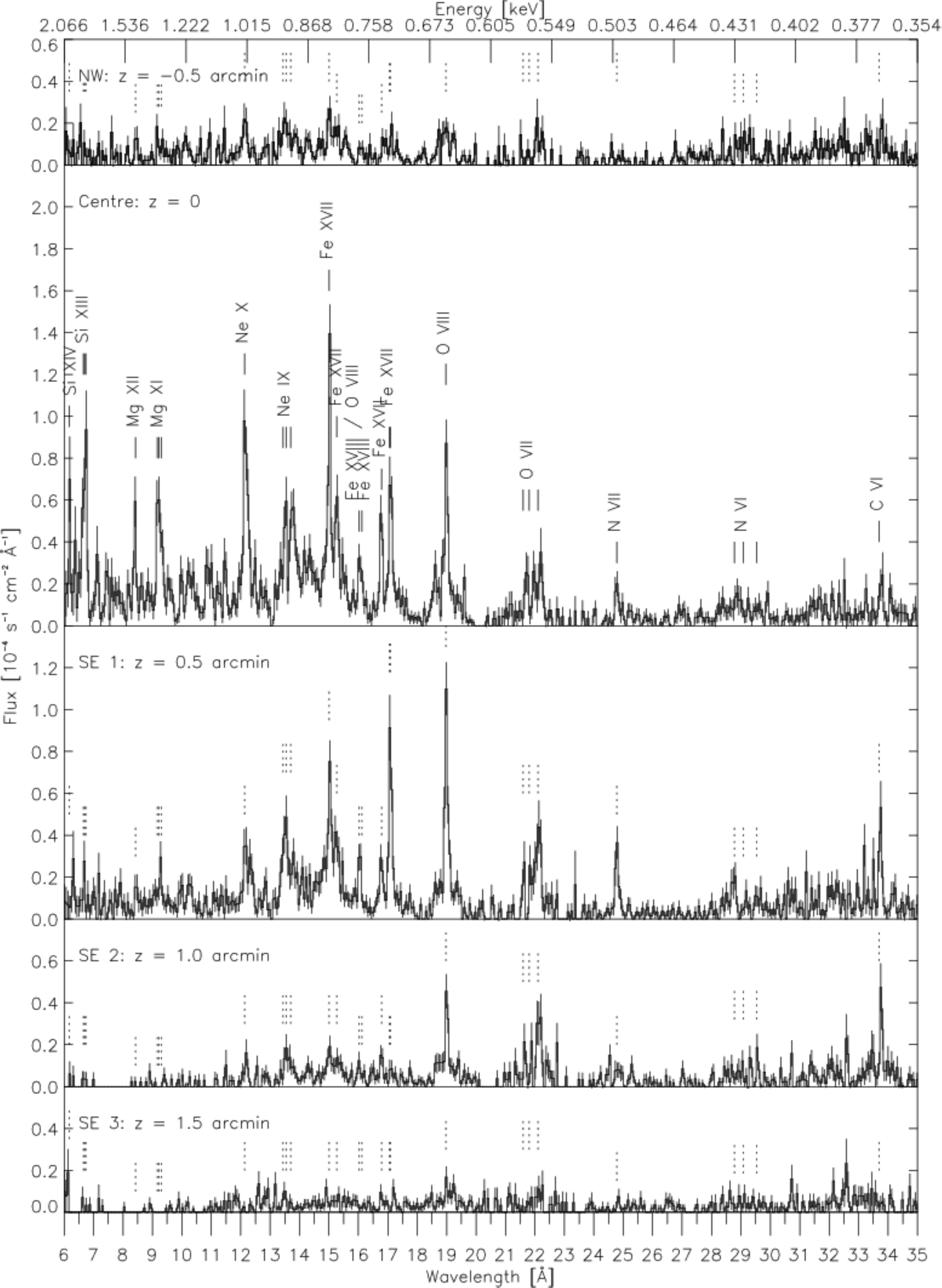}
\caption{High-resolution, soft X-ray spectra of the hot ISM in NGC\,253, obtained from different regions along the minor axis with the Reflection Grating Spectrometer (RGS) on-board \xmm (original figure from \citealt{Bauer+07}, reproduced with permission from Astronomy \& Astrophysics, $\copyright$\,ESO). The parameter `z' indicates the projected angular distance from the galaxy center (the physical scale is $\sim$1 kpc/arcmin). Note the progressive drop in emissivity moving away from the nuclear, denser region. The inherently photon-starving nature of grating spectroscopy makes these data absolutely unique: with the current facilities, such a quality can only be obtained for NGC\,253 and M\,82.}
\label{fig.02}
\end{figure*}
%
The soft X-ray emission from starburst winds is typically consistent with one to three thermal components with temperatures in the range $\sim$0.2--0.7 keV, which mimic a likely more complex temperature distribution. The X-ray spectrum of an optically-thin, hot plasma with $kT \sim 10^5$--10$^7$ K consists of a moltitude of emission lines (Figure \ref{fig.02}), which arise from atomic transitions in collisionally-excited ions and represent the most efficient channel of radiative cooling. The continuum, produced by free--free (bremsstrahlung) and free--bound (recombination) transitions, only dominates at low metallicities and high temperatures \citep{Boehringer-Hensler_89}. The X-ray spectrum of the hot ISM thus provides information not only on the gas temperature and density, which respectively govern spectral shape and overall luminosity, but also on metallicity, through the strength of the main emission lines. 

Figure \ref{fig.02} shows the high-resolution X-ray spectra of NGC\,253 obtained at different positions along the wind. All the most abundant elements usually accessible in the soft X-rays are present: oxygen, neon, magnesium, silicon (collectively referred to as $\alpha$-elements, since they are produced in the nuclear reaction sequence known as `alpha process'), and iron. As the emission lines of interest are affected by intrinsic blending and/or instrumental broadening (which becomes severe at CCD resolution), an accurate determination of the underlying continuum is mandatory for a reliable evaluation of metal abundances. This, is turn, requires a correct physical interpretation of the whole spectrum. It has been argued that integrating the emission from regions with inhomogeneous properties, owing to, for instance, temperature gradients, delivers meaningless abundance values. Such limitation plagued most of the early enrichment studies, until \chandra enabled a spatially-resolved investigation of the hot ISM properties (see, for instance, the analysis of local abundances in NGC\,4038/4039, the Antennae Galaxies, by \citealt{Baldi+06a,Baldi+06b}). In general, however, high-quality data and adequate spectral and spatial resolution are needed to safely derive elemental abundances through X-ray spectroscopy. 

With these caveats in minds, the measurement of abundances and of their ratios can be directly related to the type of SNe that control the hot ISM enrichment. This is a consequence of the markedly different delay time distribution and chemical yields of thermonuclear (Type Ia) and core-collapse (Type II, Ib/c) SNe \citep[e.g.,][]{Maoz-Graur_17}. The time delay distribution reflects the interval between the explosion of a given SN and the onset of the parent star-formation event. As the lifetime of stars with initial mass $\gtrsim 8~\msun$ is at most a few tens of Myr, virtually all core-collapse SNe will explode while the starburst episode is still in progress, so dominating the release and dispersal of metals on short timescales. Conversely, Type Ia SNe occur on average at much later times, following the evolution of the companion of the white dwarf. The delay time distribution thus depends on the shape of the initial mass function (IMF, namely the initial distribution of masses of a stellar population), and so do the various SN yields. Iron is almost equally produced by Type Ia and core-collapse SNe, as the former have roughly a ten times larger yield (0.7 vs. 0.07 $\msun$ of Fe per SN) but proportionally smaller rates (10$^{-3}$ vs. 10$^{-2}$ SNe per $\msun$ of stellar mass formed). The bulk of $\alpha$-elements is instead synthesized by core-collapse SNe. Accordingly, the $\alpha$ to Fe abundance ratios in the hot ISM of starburst galaxies are expected to be supersolar or, in the standard notation, $[\alpha/{\rm Fe}] > 0$ (where the brackets indicate that the scale is logarithmic and normalized to solar units). The value of $[{\rm O}/{\rm Fe}]$ in the halo of edge-on, star-forming disk galaxies is indeed $\sim$0.4 \citep{Strickland+04}, although iron can be partly depleted onto the dust grains carried by the wind (see Figure \ref{fig.01}). Likewise, \citet{Grimes+05} found $[\alpha/{\rm Fe}] \sim 0.5$, independent on SFR over four orders of magnitude in X-ray luminosity, in a sample that also included dwarf starbursts and ultraluminous infrared galaxies.

Superwinds are not only relevant to the chemical evolution of galaxy halos. As the outflow and/or thermal velocity of the hot gas can exceed the escape velocity of the system, a non-negligible fraction of the metals produced by the starburst will contribute to the enrichment of the inter-galactic medium (IGM). Such an efficient means of mass transport is required, for instance, by the metallicity of the intra-cluster medium (ICM), which already contained as much iron as the cluster galaxies several Gyr ago, and even at redshift $z \sim 2$ is far from being primordial in composition \citep[e.g.,][]{Mantz+18}. The maximum mass that can be lost to the IGM through a superwind can be derived from Equation \ref{eq.03} by assuming a constant SFR for 10$^7$ yr (see \citealt{Heckman+90}):
\begin{equation}
\label{eq.07}
M_{\rm esc} < 4 \times 10^7 ({\rm SFR}/\msun\,{\rm yr}^{-1})\,(v_{\rm esc}/500~{\rm km\,s}^{-1})^{-2}\quad \msun.
\end{equation}
In the most extreme cases, the entire hot ISM can be blown out of the galaxy. Unsurprisingly, Equation \ref{eq.07} also entails that the starburst-processed, superwind material is more likely to remain gravitationally bound in more massive galaxies (where the escape velocity is higher). This is a natural starting point towards the understanding of the so-called mass--metallicity relation \citep{Tremonti+04}, i.e., the tight correlation between stellar mass and gas-phase metallicity (as probed by the abundance of oxygen relative to hydrogen) observed in star-forming galaxies.

An even more obvious consequence of superwinds is the self-limiting effect on the starburst activity. Once the original gas reservoir has been exhausted, and any potential replenishment prevented (by either mechanical removal or radiative heating), star formation will rapidly subside. This negative feedback is a critical ingredient in the interpretation of the galaxy luminosity function \citep{Benson+03} and bimodality \citep{Strateva+01}. The baryonic content of galaxies is well below the cosmological value (i.e., $\sim$1/6 of the dark matter content) at all masses. Feedback is a suitable solution to the `missing baryons' problem \citep{Cen-Ostriker_06}, although in high-mass systems also an AGN contribution must be invoked (see Equation \ref{eq.07}). Bimodality arises from the peculiar distribution of the main galaxy properties, such as color, morphology, and SFR. Most galaxies fall in two distinct classes: blue, late-type, star-forming against red, early-type, passive. Again, feedback can explain the relation between galaxy color and SFR, but an additional evolutionary step must be involved to also account for the different morphologies: galaxy mergers.

\subsection{Starbursts in Galaxy Mergers}
\label{sfg-4}
As mentioned above, the highest levels of star formation or, equivalently, the shortest gas depletion times, are typically found in interacting systems. Most of the prototypical starbursts in the local Universe belong to galaxy groups, and are subject to dynamical and gravitational perturbations. This is also the case for both M\,82 \citep{Yun+94} and NGC\,253 \citep{Davidge_10}. Close interactions and mergers coincide with the most dramatic and transformational phases in the evolution of galaxies. Numerical simulations show that the tidal forces at work in a major merger between two gas-rich spirals cause a global redistribution of the cold gas, which loses angular momentum and funnels down through the potential well, so powering star formation at starburst rates and possibly SMBH accretion \citep{Mihos-Hernquist_96}. The end product of this process is believed to be a massive, quiescent elliptical on both theoretical and observational grounds \citep{Hopkins+06,Dasyra+06}.

Simulations of galaxy mergers also suggest that the shocks associated with the collision can produce significant amounts of X-ray emitting gas, whose temperature and luminosity strongly depend on the properties of the progenitors (mass, gas fraction, metallicity) and on the orbital parameters of the collision (relative orientation, angular momentum). A different time evolution is predicted based on whether galactic disks or halos are considered, whereby the peak X-ray luminosity is reached at the final coalescence or the first perigalactic passage, respectively \citep{Cox+06,Sinha-HolleyB_09}. However, none of these models properly include the role of starburst-driven shocks, which are actually responsible for the dominant heating effects. In fact, the observations of equal-mass galaxy pairs caught at different epochs along the merger sequence indicate that the ratio between thermal X-ray luminosity of the hot ISM and SFR is nearly constant (with an rms scatter of $\approx$0.3 dex; \citealt{Smith+18}) irrespective of the merger stage, at:
\begin{equation}
\label{eq.08}
L_{\rm X,GAS}/{\rm SFR} \sim 5.5 \times 10^{39} ~ {\rm erg\ s}^{-1}\, (\msun\ {\rm yr}^{-1})^{-1}. 
\end{equation}
This implies that {\it (i)} the mechanical energy of starburst winds is converted into X-ray radiation with a typical efficiency of a few per cent (cf. Equation \ref{eq.03}); and {\it (ii)} the diffuse X-ray emission must trace, almost instantaneously, the star formation history during the merger. 

It should be noted that merger-induced SMBH accretion is a competitive mechanism to heat and disperse the gas. For an AGN with bolometric luminosity $L_{\rm bol} \sim 10^{45}$ erg s$^{-1}$, and a coupling efficiency between radiation field and ISM of $\approx$1\%, the energy input is commensurate with that from the starburst, and capable of initiating a large-scale feedback process \citep[e.g.,][]{Hopkins-Elvis_10}. Yet, the actual AGN contribution is expected to be minor until after the coalescence, as models and observations confirm that nuclear activity is heavily obscured at earlier times. Eventually, accretion becomes very efficient (i.e., Eddington-limited), the black hole quickly gains mass, and the resulting feedback regulates its further growth and quenches the global star formation by removing the gas supply from the inner regions \citep{Springel+05,Hopkins+05}. The presence of a powerful AGN and its complex interplay with the starburst therefore determine how rapidly the X-ray luminosity fades and how much of the hot gas remains bound to the system after the merger (see also Section \ref{etg-p.12}).

\begin{figure*}[t!]
\centering
\includegraphics[width=0.85\textwidth]{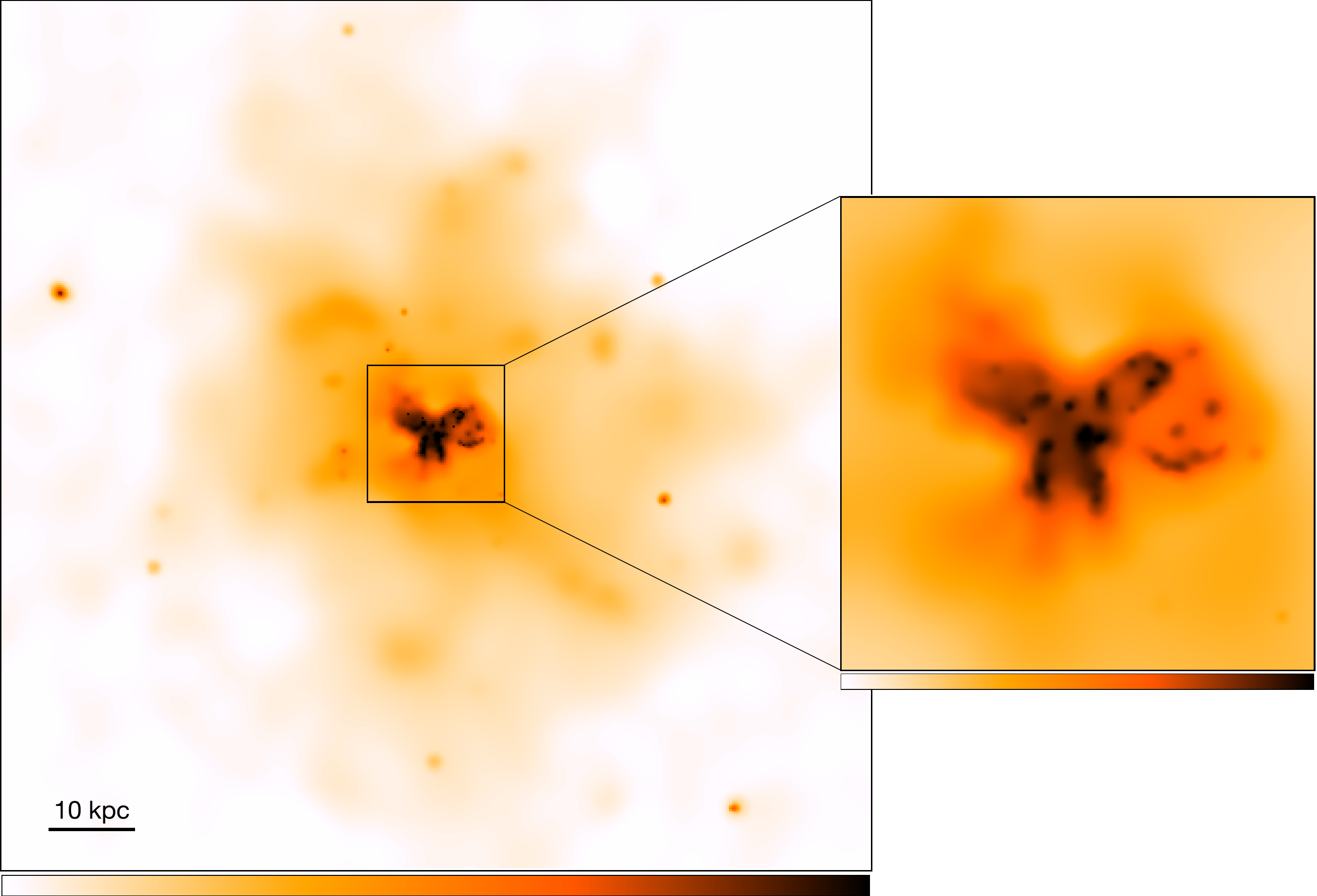}
\caption{Soft X-ray (0.5--1.5 keV) image of NGC\,6240, obtained by co-adding 182 ks of \chandra exposures. The 100-kpc wide field of view is almost entirely filled by the hot halo, which presents several substructures consistent with widespread superwind activity. The zoom-in on the right highlights the complex morphology of the central, butterfly-shaped nebula, powered by the most recent starburst episode. Both images were background-subtracted and adaptively smoothed with Gaussian kernels. The dynamic range of the color bar is 1,200 for the halo and 2,400 for the nebula, respectively, and the scale is logarithmic. Adapted from \citet{Nardini+13}.}
\label{fig.03}
\end{figure*}

\subsection{An Ideal Laboratory: NGC\,6240}
\label{sfg-5}
Deep X-ray observation of galaxy mergers are rather scarce, but they invariably reveal a wealth of details. One of the most spectacular examples is represented by NGC\,6240 ($D \sim 107$ Mpc), an IR-luminous, morphologically disturbed galaxy pair on its way to entering the final coalescence stage (see \citealt{Nardini+13}, and references therein). As NGC\,6240 typifies in a single object the broad phenomenology of galaxy mergers, and it can be regarded as an ideal test bed for the predictions of numerical models, it is worth discussing its properties at some length for illustrative purposes. 

The central kpc of NGC\,6240 harbours two AGNs, still buried inside the remnants of the former galactic bulges, and $\sim$10$^{10}~\msun$ of molecular gas, ruffled by turbulence, shocks, and outflows. A $\sim$10-kpc wide, butterfly-shaped nebula, characterized by clumps, loops, and filaments seen in both soft X-rays (Figure \ref{fig.03}, blow-up) and H$\alpha$, is the telltale signature of the action of superwinds, driven by a nuclear starburst that forms several tens of $\msun$ yr$^{-1}$. The entire system is surrounded by a huge halo of hot ($\sim$0.65 keV, or 7.5 million K) gas, with an average radial extent of $R_{\rm halo} \sim 50$ kpc, which accounts for about one third of the total diffuse X-ray emission (Figure \ref{fig.03}). Its luminosity exceeds $4 \times 10^{41}$ erg s$^{-1}$, and is more typical of galaxy groups \citep{Mulchaey_00} and massive ellipticals \citep{Mathews-Brighenti_03}. Interestingly, the possible fossil group nature of NGC\,6240 is supported by the recent discovery of a third nucleus in the previously unresolved southern component \citep{Kollatschny+20}, while the ultimate fate of the merger is anticipated by its stellar surface brightness, which is already conforming to the $\propto r^{1/4}$ radial profile characteristic of spheroids \citep{Bush+08}. 

It can be easily demonstrated that the inelastic nature of the collision is not sufficient in itself to account for the thermal energy content of the halo. For simplicity, let two identical progenitors collide head-on with relative speed $v_{\rm coll}$. The kinetic energy dissipated in the merger is then approximately $M_{\rm X,GAS} v_{\rm coll}^2 /8$, where $M_{\rm X,GAS}=n_e \eta V m_p$ is the total mass of X-ray emitting gas. As the thermal energy of the halo is $E_{\rm th}=3n_e \eta V kT$, it immediately follows that a collision velocity of $v_{\rm coll} \sim 1,200$ \kms is required to heat the gas up to $kT \simeq 0.65$ keV, far too high for any known galaxy merger. Also gravitational confinement, which is an important heating source for the halos of virialized systems (Sections \ref{etg-k.3} and \ref{etg-p.12}), does not seem a viable explanation for NGC\,6240. Indeed, by using the gas velocity dispersion $\langle\sigma\rangle$ as an equivalent probe of the depth of the potential well,\footnote{Note, however, that this strictly applies only for systems in hydrostatic equilibrium (Section \ref{etg-p.4}).} one gets a virial temperature $kT_{\rm VIR} \sim \mu m_p \langle\sigma\rangle^2$ of only $\sim$0.25 keV for the average value of $\langle\sigma\rangle \sim 200$ \kms found in the core \citep{Treister+20}.

It is rather straightforward to ascribe the gas heating to starburst-driven winds instead. The kinetic luminosity of a starburst, i.e., the mechanical energy injected into the ISM per unit time ($\sim \xi {\dot E_{\rm SB}}$; Equation \ref{eq.03}), is roughly 1\% of its bolometric luminosity \citep{Leitherer+99}. This is $L_{\rm kin} \sim 3 \times 10^{43}$ erg s$^{-1}$ in NGC\,6240, to be compared to the thermal energy of the halo, $E_{\rm th} \simeq 5 \times 10^{58} \eta^{1/2}$ erg. Hence, it takes less than 50 Myr for the starburst to supply the required energy, as opposed to a dynamical age of the halo, $2R_{\rm halo}/c_s$ (where $c_s$ is the adiabatic sound speed), of over 200 Myr. Given the halo size, however, a wind from a centrally localized starburst should have traveled at $v_{\rm wind} \gtrsim$ 1,000 \kms, which is unlikely for dense environments like the nuclear regions of galaxy mergers (where $\beta \gg 1$). Also, the mild decline with radius of the X-ray surface brightness is not compatible with a free-flowing wind. It seems then plausible that the hot halo stems from the superposition of successive winds, emanating from different locations and escaping in different directions over the last 50--200 Myr. 

As the physical properties of the halo do not seem to vary significantly with either radius or position angle, in keeping with the proposed `widespread wind' scenario, the single-temperature, $kT \simeq 0.65$ keV model provides a reasonable first approximation of the basic properties of the hot gas. This notwithstanding, the spectrum of the full halo is best described by two gas phases with different temperatures and metallicities: a `hot, metal-rich' component, with $kT \sim 0.8$ keV, $Z_\alpha$ consistent with (or slightly below) solar, and $[\alpha/{\rm Fe}] \approx 0.6$; and a `warm, metal-poor' component, with $kT \sim 0.25$ keV and $Z \sim 0.1$ \zsun. It is therefore tempting to identify the hot phase with the chemically evolved, starburst-injected gas, which is dispersing into a pristine, pre-existing ambient medium, heated itself to X-ray emitting temperatures by gravitational infall and/or merger-related dissipation (see above). As a matter of fact, this might still be an oversimplification, but the current data do not allow any more exhaustive analysis.

To date, only a handful of X-ray observations exist with comparable quality to that of NGC\,6240. Once some obvious differences (e.g., mass and gas fraction of the progenitors, merger stage) are considered, the general picture drawn from the case study of NGC\,6240---widespread starburst winds and redistribution of metals across the halo---is broadly confirmed \citep{Veilleux+14,Liu+19}. This holds true even after moving to relatively larger samples, at the cost of shallower exposures. Most of the advanced mergers exhibit diffuse X-ray emission beyond a distance of 10 kpc \citep{Huo+04}. The metallicity of the gas in these extended halos is tentatively constrained to be significantly subsolar ($Z \approx 0.1$ \zsun), as if there had not yet been any pollution from the central starbursts. This is not unexpected, as mass loading is much more severe in mergers than in M\,82-like objects, and this can substantially slow down the wind. More reliable measurements of both temperature and metallicity out the largest possible scales are needed to understand whether this pristine gas already belonged to the colliding galaxies or is accreted from the IGM, and how much of the hot ISM will be retained to form the halo of a young elliptical galaxy once the merger is over. 

\section{Observational Properties of the Hot ISM in Early-Type Galaxies}
\label{etg-k}
The hot ISM is the dominant phase of the ISM in early-type galaxies (ETGs). Its physical and chemical properties are shaped by the processes occurring during the formation and evolution of the host galaxy. By measuring the current status of the hot ISM, we can then infer how critical astrophysical processes worked throughout the galaxy's history. These processes include AGN feedback, environmental effects (e.g., merger, accretion, and stripping), and star formation and its quenching (e.g., \citealt{Kim-Pellegrini_12}, and references therein; see also Section \ref{sfg}). With state-of-art X-ray observations, we can measure the physical and chemical properties of the hot ISM. These measurements put constraints on the global ISM properties, such as X-ray luminosity ($L_{\rm{X,GAS}}$), temperature ($T_{\rm X}$), and also chemical abundances (primarily Fe; Section \ref{sfg-3}). We can also study the spatial distributions of these properties and of the physical parameters that we can derive from them: density, entropy, pressure, mass of the hot ISM, and total galaxy mass. The observational constraints on these quantities can then be compared with theoretical predictions. In this Section, we will describe what we have learned from the X-ray observations of hot ISM in ETGs.

\subsection{From Discovery with the \einstein Observatory to \chandra and \xmm}
\label{etg-k.1}
Since the launch of the \einstein Observatory in 1978, X-ray bright, giant ETGs have been known to host large amounts of hot gas (e.g., \citealt{Forman+85,Trinchieri-Fabbiano_85}). Early studies could only measure the integrated X-ray luminosity and temperature of ETGs ($L_{\rm{X,TOT}}$, and $T_{\rm TOT}$), and these measurements were used to constrain the properties of the hot ISM (see the review by \citealt{Fabbiano_89}). However, the X-ray emission is coming not only from the X-ray emitting, hot gas, but also from X-ray binaries---mostly low-mass X-ray binaries (LMXBs) in the old stellar populations of ETGs \citep{Trinchieri-Fabbiano_85}. Moreover, active nuclei within the galaxy may also contribute to the X-ray emission, and the X-ray emission of background galaxies and AGNs may contaminate the data. While $L_{\rm{X,TOT}}$ is close to $L_{\rm{X,GAS}}$ in gas-rich, X-ray luminous ETGs, there are also gas-poor, X-ray faint ETGs where $L_{\rm{X,TOT}}$ may be dominated by LMXBs \citep{Trinchieri-Fabbiano_85}. Some ETGs may also host an AGN.

Launched in 1999, the \chandra X-ray Observatory, with its sub-arcsecond spatial resolution, has revolutionized our understanding of the X-ray emission of ETGs. Typical \chandra observations of ETGs at distances of 10--20 Mpc, with an exposure of several hours can detect some 100 point-like sources in an individual galaxy (e.g., NGC\,1399, \citealt{Angelini+01}; NGC\,4365 and NGC\,4382, \citealt{Sivakoff+03}; see also \citealt{Kim-Fabbiano_04a} and the review by \citealt{Fabbiano_06}). These observations have led to a good characterization of the X-ray luminosity functions of LMXBs in different environments, e.g., in the field versus globular clusters (GCs), metal-rich versus metal-poor systems, and young versus old stellar populations (e.g., Figure 4 in \citealt{Kim-Fabbiano_10}, and references therein). The capability of detecting individual point sources is also valuable to exclude contaminating sources for accurate measurements of the hot gas properties. Thanks to \chandra, we can, for the first time, effectively separate the different X-ray emission components of ETGs and accurately measure the properties of the hot gas, $L_{\rm{X,GAS}}$ and $T_{\rm X}$ (e.g., \citealt{Boroson+11,Kim+19b}).

With its large field of view and higher sensitivity, \xmm, also launched in 1999, provides complementary observational data. \xmm data are particularly useful in studying the faint diffuse emission in the outskirts of ETGs and in measuring chemical properties (e.g., Fe abundance) that often require high S/N ratio spectra (e.g., \citealt{Islam+21}).

\subsection{Global Properties of the Hot ISM: Scaling Laws}
\label{etg-k.2}
X-ray scaling relations have been built starting from the \einstein X-ray mission based on the sample of observed ETGs, and widely used to compare observations with model predictions. More recently, following the realization of the existence of two main families of ETGs, different X-ray scaling relations were derived for each of these ETG families: {\it (i)} core galaxies, with a flattened central surface brightness distribution, which typically have large stellar masses, high stellar velocity dispersion, round isophotes, old stellar populations, and slowly rotating stellar kinematics; and {\it (ii)} power-law (also called coreless) galaxies, which instead have a centrally rising surface brightness distribution, tend to be smaller, disky, fast rotating, and with some recent star formation (see \citealt{Pellegrini_05,Kormendy+09,Sarzi+13}). \citet{Kormendy+09} suggested that core ETGs are primarily formed by dry mergers, while coreless ETGs may be the product of gas-rich, wet mergers, with an ensuing period of intense stellar formation (see also \citealt{Binney_04,Nipoti-Binney_07}). The X-ray scaling relations were separately built for core and coreless galaxies to understand their distinct characteristics \citep{Kim-Fabbiano_15}, and further compared with other systems (groups and clusters of galaxies) and simulations to constrain various model parameters (e.g., \citealt{Negri+14a,Negri+14b,Negri+15}). 

The most used scaling relation in the early days, the $\lx$--$L_B$ relation, compared the total X-ray and optical ($B$ band) luminosities of ETGs. This relation, which showed a factor of 100 scatter in $L_{\rm{X,TOT}}$ for a given $L_B$, triggered a series of theoretical investigations (e.g., \citealt{Fabbiano_89}, and references therein). Thanks to the resolution of \chandra, $L_{\rm{X,GAS}}$ is now used in place of $L_{\rm{X,TOT}}$. In the $L_{\rm{X,GAS}}$--$L_B$ (or $L_{\rm{X,GAS}}$--$L_K$, see Section \ref{etg-p.11}, Figure \ref{fig.14}) relation, the scatter is even more significant, a factor of 1,000 \citep{Kim-Fabbiano_15}. This is because X-ray faint ETGs hold only a small amount of hot gas; therefore, $L_{\rm{X,GAS}}$ may be considerably less than $L_{\rm{X,LMXB}}$ \citep{Boroson+11}. The critical astrophysical factors that govern the amount of hot gas in ETGs are discussed in Section \ref{etg-p.1}.

The temperature of the hot ISM ($T_{\rm X}$), instead, directly reflects the potential depth of a virialized system. Furthermore, $L_{\rm{X,GAS}}$ and $T_{\rm X}$ can be measured with a single X-ray observation, but independent methods: $L_{\rm{X,GAS}}$ with wide-band photometry (or normalization of the best-fit model); $T_{\rm X}$ with model fits of observed spectral data. Once the gas contribution has been `cleaned' of contaminants (e.g., LMXBs) in the \chandra data, the $L_{\rm{X,GAS}}$--$T_{\rm X}$ relation could be reliably built \citep{Kim-Fabbiano_15,Goulding+16} and extensively compared with model predictions \citep{Choi+15,Choi+17,Ciotti+17}.

\begin{SCfigure}[][t!]
\includegraphics[width=0.55\textwidth]{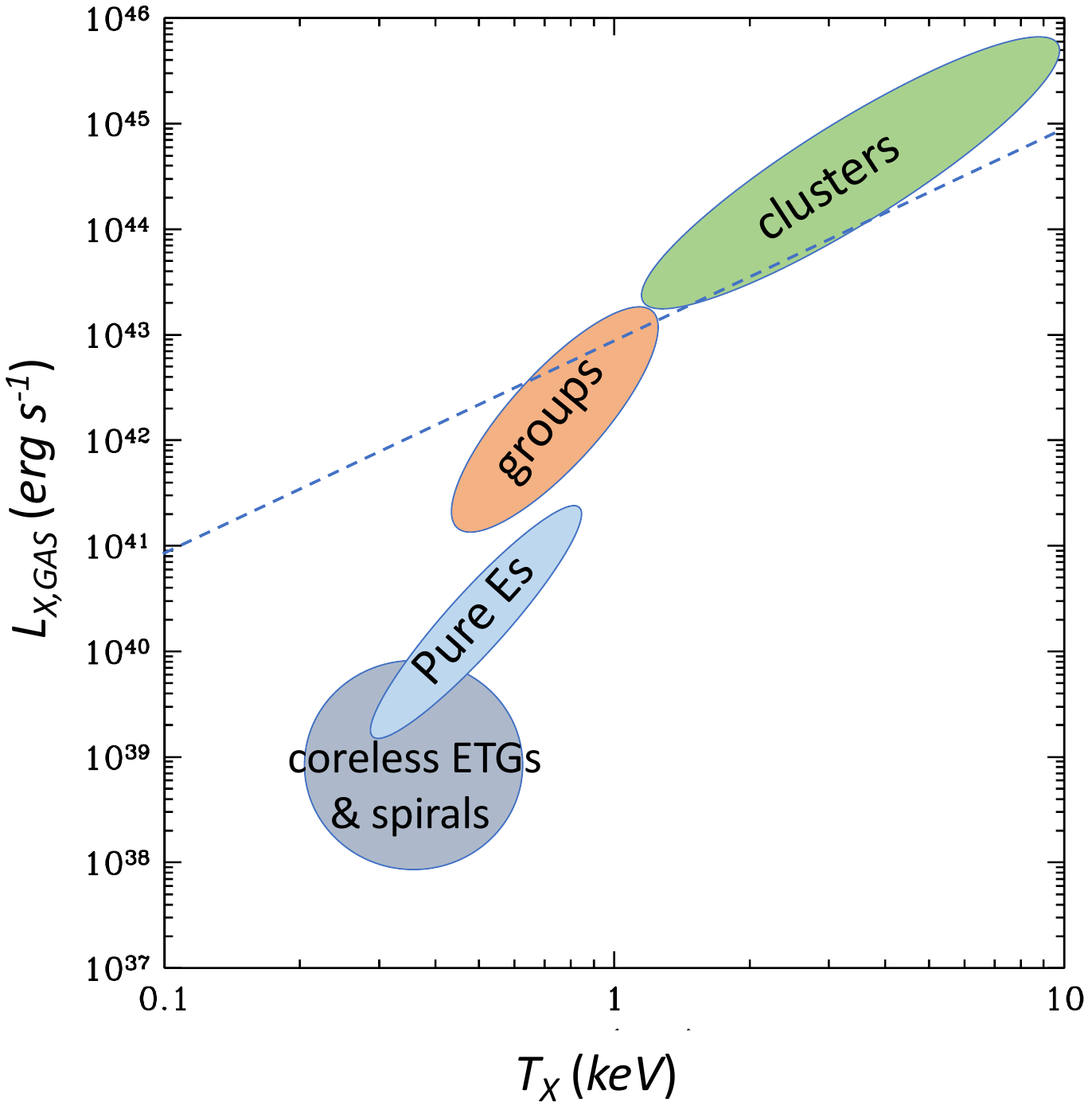}
\caption{Comparison of the $L_{\rm{X,GAS}}$--$T_{\rm X}$ relations in various samples, taken from \citet{Kim-Fabbiano_15}. From the bottom left, coreless ETGs and spirals have no correlation, while core (`pure') E galaxies have a very tight correlation, $L_{\rm{X,GAS}} \propto T_{\rm X}^{4.5}$. Groups have a similar trend as the core Es, but they are shifted toward higher $L_{\rm{X,GAS}}$. Clusters at the top right corner have a flatter relation ($L_{\rm{X,GAS}} \propto T_{\rm X}^3$), compared to the other sub-samples. For a reference, the flatter self-similar expectation ($L_{\rm{X,GAS}} \propto T_{\rm X}^2$) is shown in dashed lines.}
\label{fig.04}
\end{SCfigure}

Figure \ref{fig.04} shows the $L_{\rm{X,GAS}}$--$T_{\rm X}$ relation of systems ranging from massive clusters to individual galaxies. For coreless ETGs (at the bottom left)---these galaxies also tend to show stellar rotation, a flattened galaxy figure, and rejuvenation of the stellar population---$L_{\rm{X,GAS}}$ and $T_{\rm X}$ are not correlated. The $L_{\rm{X,GAS}}$--$T_{\rm X}$ distribution of coreless ETGs is a scatter diagram clustered at $L_{\rm{X,GAS}} < 10^{40}$ erg s$^{-1}$, similar to that reported for the hot ISM of spiral galaxies (e.g., \citealt{Li-Wang_13}), suggesting that both the energy input from star formation and the effect of galactic rotation and flattening may disrupt the hot ISM.

The relation is steep and tight for core ETGs ($L_{\rm{X,GAS}} \propto T_{\rm X}^{4.5}$). At the high luminosity end, the $L_{\rm{X,GAS}}$--$T_{\rm X}$ correlation is similar to that found in samples of groups but shifted down toward relatively lower $L_{\rm{X,GAS}}$ for a given $T_{\rm X}$. This relation is considerably steeper than both the relation for clusters ($L_{\rm{X,GAS}} \propto T_{\rm X}^3$; \citealt{Arnaud-Evrard_99}) and the expectation from the self-similar model, where gravity dominates ($L_{\rm{X,GAS}} \propto T_{\rm X}^2$, blue dashed line in Figure \ref{fig.04}). The slope of $\sim$3\ in clusters, steeper than the self-similar case, indicates that baryonic physics is already important even on this large scale. The fact that the slope in core ETGs is even steeper ($\sim$4.5) further demonstrates the increase of importance of non-gravitational effects, including AGN and stellar feedback. Therefore, the overall trend may be understood by considering the relative importance of baryonic physics over the pure gravity in the galaxy scale via AGN and stellar feedback (for more discussion see \citealt{Kim_17}).

\begin{SCfigure}[][t!]
\includegraphics[width=0.55\textwidth]{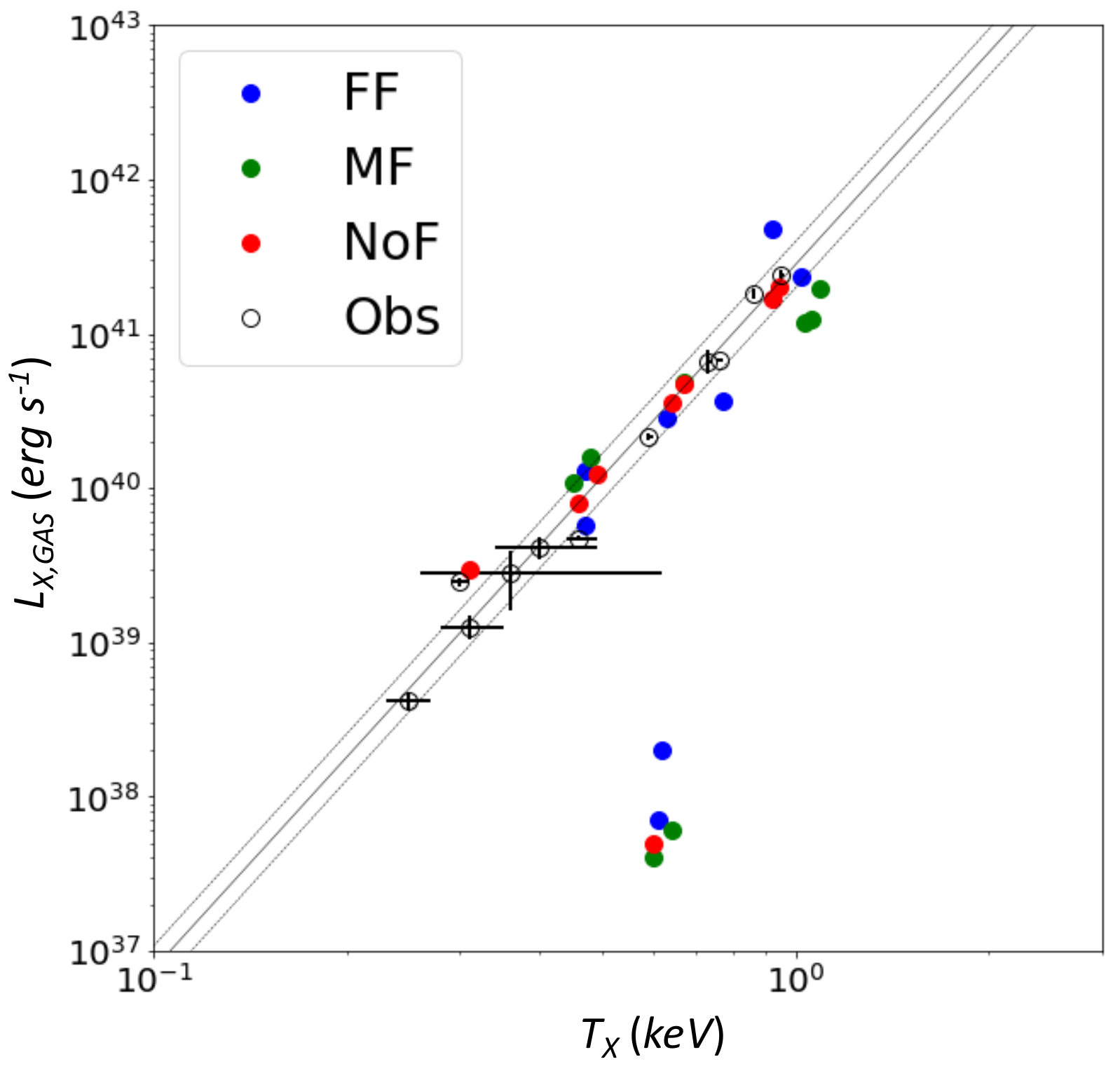}
\caption{$L_{\rm{X,GAS}}$--$T_{\rm X}$ relation for an E sample (black open circles), taken from \citet{Kim-Fabbiano_15}. The diagonal lines are the best fit (solid line) with rms deviations (dashed lines). The predictions from equivalent (i.e., low-rotation) models with three different AGN feedback recipes---full (FF: blue), mechanical (MF: green), no (NoF: red) feedback---are also plotted for comparison (taken from \citealt{Ciotti+17}).}
\label{fig.05}
\end{SCfigure}

Figure \ref{fig.05} shows a zoom-in view of core ETGs with the results of observations and simulations, from \citet{Kim-Fabbiano_15}. Only core elliptical (E) galaxies with no sign of recent star formation are plotted (open black circles with error bars). This relation is tight, with rms deviation of only 0.13 dex. For $L_{\rm{X,GAS}} > 10^{40}$ erg s$^{-1}$, this correlation compares exceptionally well with the predictions of high-resolution hydrodynamical simulations for fully velocity-dispersion-supported galaxies (the model predictions are taken from \citealt{Ciotti+17}, their Fig. 2; see also \citealt{Negri+14a,Negri+14b,Choi+15,Choi+17}. Simulations and model specifications are further discussed in Sections \ref{etg-p.10} and \ref{etg-p.12}).

However, the observed correlation extends down to $L_{\rm{X,GAS}} \sim 10^{38.5}$ erg s$^{-1}$, contrary to the simulations, which predict a sudden drop in X-ray luminosity in cooler galaxies. 
The models predict considerably lower $L_{\rm{X,GAS}}$, or higher $T_{\rm X}$ (blue and green circles at the middle bottom). This can be understood because the hot ISM in small systems is in the outflow/wind state, and then it tends to be at higher $T_{\rm X}$ than extrapolated from the $L_{\rm{X,GAS}}$--$T_{\rm X}$ relation of the inflow state in large systems (e.g., \citealt{Pellegrini_11,Negri+14a,Negri+14b}). 
Such a discrepancy is yet to be understood, possibly suggesting a need to improve feedback recipes (see also Section \ref{etg-p.12}).

\subsection{The Mass of ETGs}
\label{etg-k.3}
The $K$-band luminosity is a good proxy for the integrated stellar mass of the galaxy ($L_K \sim M_\star$ in solar units; \citealt{Bell+03}), but it does not measure the amount of dark matter (DM), which may be prevalent especially at large radii. The total mass ($M_{\rm TOT}=M_\star+M_{\rm DM}$), out to radii comparable to the full extent of the hot halos of gas-rich ETGs, is the crucial quantity to effectively explore the importance of gravitational confinement for the hot gas retention (see \citealt{Mathews+06}). Note that the amount of gas mass itself is small in ETGs (unlike large-scale clusters), and not important for gravitational confinement (e.g., \citealt{Canizares+87}).

Although $M_{\rm TOT}$ could be measured with X-ray observations (e.g., \citealt{Mathews+06}) under the assumption that the hot gas is in hydrostatic equilibrium, the X-ray measurements suffer from a couple of issues: {\it (i)} this method is only applicable to hot, gas-rich systems; {\it (ii)} the hot ISM may deviate from hydrostatic equilibrium, as revealed by observed dynamical evidence (e.g., sloshing, cavities; see \citealt{Kim+19a}), and the X-ray measurements are sometimes appreciably different from other independent results (e.g., \citealt{Paggi+17}). These issues may be less significant in relaxed clusters of galaxies, but they are not negligible in typical ETGs. 

Dynamical masses have been measured for a large number of ETGs using integral field 2D spectroscopic data (e.g., Atlas$^{\rm 3D}$, \citealt{Cappellari+13}; MaNGA, \citealt{Li+18}). However, these data are limited to radii within $r < 0.5$--1 $R_{\rm e}$ (where $R_{\rm e}$ is the effective, or half-light, radius), considerably smaller than the observed extent of the hot gas in gas-rich ETGs. Strong gravitational lenses also provide lensing masses (e.g., \citealt{Auger+10}), but the measurements are again limited to $r < 0.5$--1 $R_{\rm e}$. A (small) number of dynamical mass measurements at large radii have recently become available from the analysis of the kinematics of hundreds of GCs and planetary nebulae (PNe) in individual galaxies \citep{Deason+12,Alabi+17}. If a hot halo is gravitationally confined, the amount of this gas, which is related to the measured X-ray luminosity, ought to be related to the total mass of the ETG (stellar and dark matter). To test this hypothesis, \citet{Kim-Fabbiano_13} compared $L_{\rm{X,GAS}}$ with independently measured (i.e., non-X-ray based) $M_{\rm TOT}$ of 14 ETGs taken from \citet{Deason+12}. \citet{Forbes+17} extended to 29 ETGs with GC-based $M_{\rm TOT}$ taken from \citet{Alabi+17}, and found indeed a good correlation.

\begin{SCfigure}[][t!]
\includegraphics[width=0.55\textwidth]{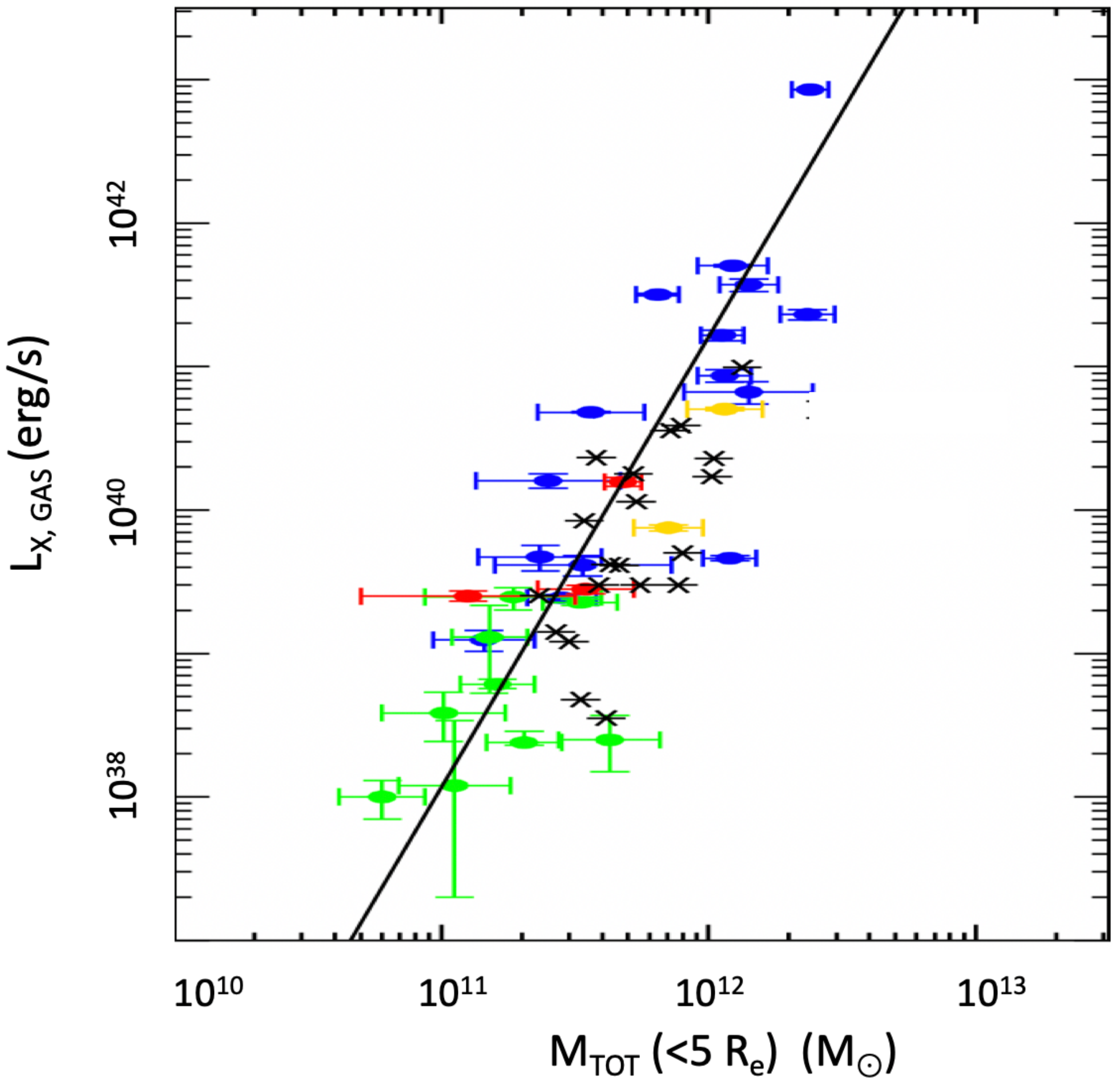}
\caption{X-ray luminosity of the hot gas versus total mass within $r < 5R_{\rm e}$, taken from \citet{Forbes+17}. ETGs are color-coded by their central optical light profile (blue: core, gold: intermediate, green: coreless, red: unknown). Model galaxies from \citet{Choi+15} are also shown as black stars.}
\label{fig.06}
\end{SCfigure}

Figure \ref{fig.06} shows the $L_{\rm{X,GAS}}$--$M_{\rm TOT}$ relation. The relation is tight, with a slope $\alpha = 3.13 \pm 0.32$ in $L_{\rm{X,GAS}} \propto M_{\rm TOT}^\alpha$, indicating that $M_{\rm TOT}$ is indeed the primary factor in regulating the amount of hot gas retained by the galaxy. This relation holds in the range of $L_{\rm{X,GAS}} = 10^{38}$--10$^{43}$ erg s$^{-1}$ (spanning five orders of magnitude), or $M_{\rm TOT} =$ a few $\times$10$^{10}$\,--\,a few $\times$10$^{12}~\msun$ (spanning two orders of magnitude). Also shown in Figure \ref{fig.06} are the results (black stars) from the cosmological simulations by \citet{Choi+15}. Although they only cover a limited range, their predictions agree well with the observations of $L_{\rm{X,GAS}}$ and $M_{\rm TOT}$ within 5$R_{\rm e}$. Separating core (blue points) and coreless (green points) ETGs, both \citet{Kim-Fabbiano_13} and \citet{Forbes+17} suggested that core ETGs reveal a tighter relation than coreless ETGs. This is consistent with the findings of \citet[][see also Figure \ref{fig.04}]{Kim-Fabbiano_15}, and suggests that other factors, such as stellar feedback, may play an increased role in coreless (low-mass) ETGs. Although $L_{\rm{X,GAS}}$ of both core (high $\lx$) and coreless (low $\lx$) ETGs are primarily determined by total mass, hence the depth of the potential well, other factors may still play a non-negligible role for galaxies with lower mass.

As a growing number of ETGs have reliable measurements of their central SMBH mass ($M_{\rm BH}$), it is interesting to compare $M_{\rm BH}$ and $L_{\rm{X,GAS}}$. This relationship provides another excellent example of how the hot gas properties can be used to address essential aspects of ETGs, i.e., how the central SMBHs, host galaxies, and DM halos co-evolve. Through numerical simulations, \citet{Booth-Schaye_10} suggested that the DM halo may determine $M_{\rm BH}$. However, \citet{Kormendy-Bender_11} presented an opposing argument that $M_{\rm BH}$ is not correlated directly with the DM halo (see also \citealt{Kormendy-Ho_13}). \citet{Bogdan-Goulding_15} investigated this issue by stacking the {\it ROSAT} all-sky survey data of the large sample of SDSS elliptical galaxies. They suggested that the central stellar velocity dispersion, hence the central gravitational potential, may be more tightly connected to the total (large-scale) mass, traced by $L_{\rm{X,GAS}}$ with the relationship in \citet[][similar to Figure \ref{fig.06}]{Kim-Fabbiano_13}, than to the stellar mass ($M_\star$). \citet{Gaspari+19} further analyzed the \chandra and \xmm data of 85 systems (including ETGs and clusters). They found that the tightest correlation is between $M_{\rm BH}$ and the hot gas properties ($L_{\rm{X,GAS}}$ and $T_{\rm X}$), instead of those determined by the optical data (e.g., $\sigma$, $M_\star$). They again suggested that the hot halos play a more central role than stars in tracing and growing SMBHs. The direct (or indirect) connection between DM halos and SMBHs, if confirmed, can help us to fully understand the evolution of ETGs through feedback mechanisms. 

\subsection{1D Radial Profiles of the X-ray Surface Brightness and Temperature Distributions}
\label{etg-k.4}
Once the radial profile of the X-ray surface brightness of diffuse hot gas has been measured and compared with the optical light profiles, it was immediately noticed that the hot gas distribution is different from the stellar distribution (e.g., \citealt{Trinchieri+86,Fabbiano+92}). The hot gas is extended with a flatter slope than the stellar light in hot-gas-rich galaxies, while it is steeper and confined to the central region in gas-poor galaxies. The hot gas-rich systems are typically giant, old (core) elliptical galaxies, often sitting at the center of DM-rich systems (e.g., NGC\,1399 at the center of the Fornax cluster). The hot gas-poor systems are relatively small, coreless, and may have some ongoing star formation (e.g., NGC\,3379, \citealt{Trinchieri+08}; NGC\,4278, \citealt{Pellegrini+12b}). 

Another critical piece of evidence that was recognized from the early observational and theoretical investigations of the X-ray surface brightness profile is the absence of a strong central peak (e.g., \citealt{Ciotti+91,Mathews-Brighenti_03}), which would result from the uncontrolled gas inflow (the cooling catastrophe) predicted by the early cooling flow models (e.g., \citealt{Fabian_94}; see Section \ref{etg-p} for the discussion of feedback mechanisms that control the gas flow).

In addition to the X-ray surface brightness profile, the radial profiles of the spectral and chemical properties of the hot gas are of interest. They include the temperature, Fe abundance and $\alpha$-elements to Fe ratios, density, pressure, and entropy. All the density-based quantities require 3D deprojection. Because of the uncertainties introduced in deprojection and limitations in data quality, the projected pressure and entropy (based on the projected surface brightness rather than the density) are sometimes used as a proxy.

\begin{figure}[t!]
\centering
\includegraphics[width=0.95\textwidth]{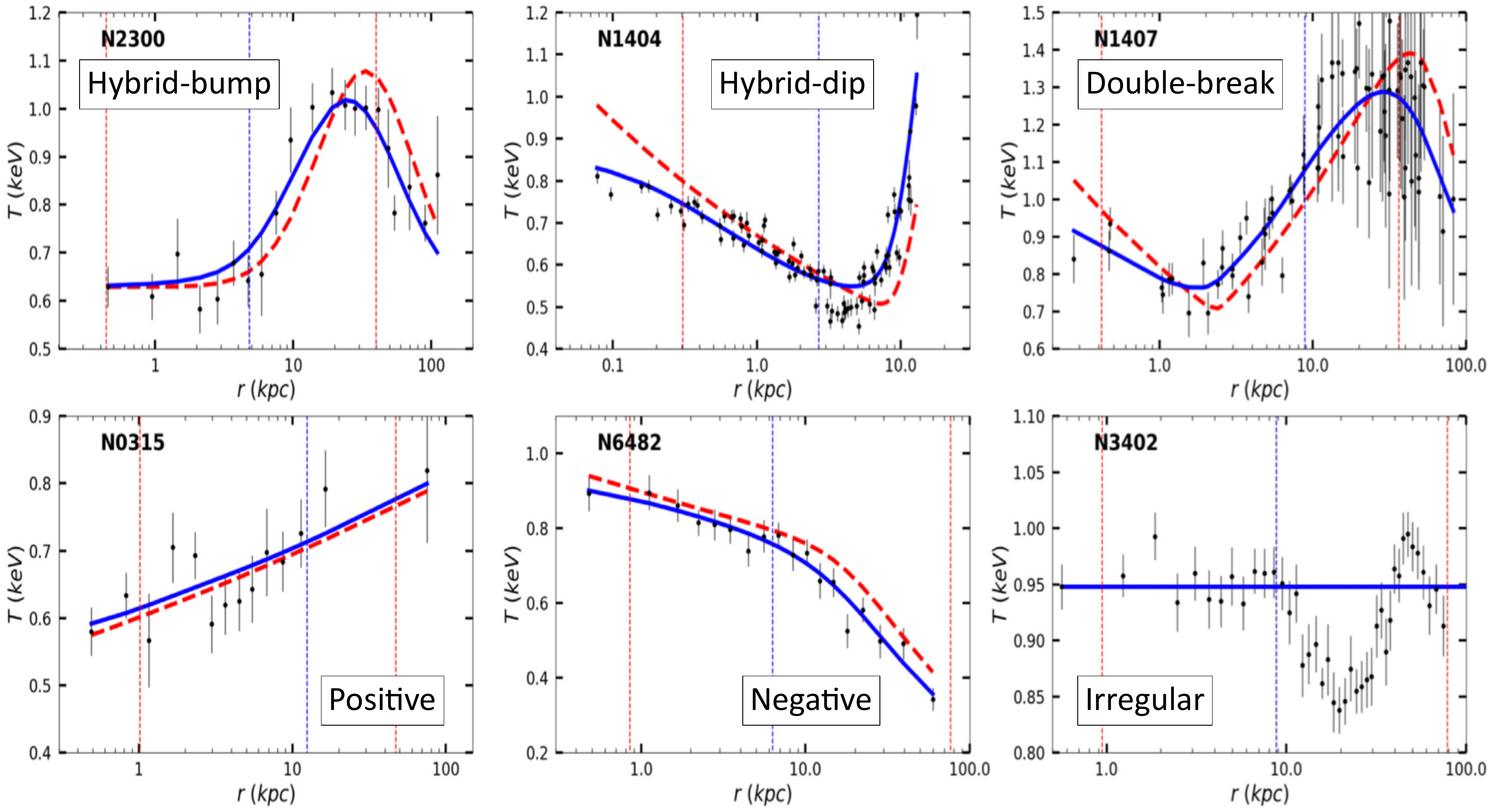}
\caption{Examples of six different temperature profile types. The type and the example galaxy are specified in each plot. The data points in black are fitted with projected temperature profiles in blue, with the 3D model shown in a red dashed line. The inner red vertical line indicates $r = 3$ arcsec, where the AGN could affect the temperature measurement, and the outer red line indicates the maximum radius where the hot gas emission is reliably detected with an azimuthal coverage larger than 95 per cent. The blue vertical line is at one effective radius. Taken from \citet{Kim+20}.}
\label{fig.07}
\end{figure}

The radial profile of the hot gas temperature $T_{\rm X}$ is shaped by various heating mechanisms, including AGN feedback \citep{Fabian_12}, stellar feedback \citep{Ciotti+91}, and gravitational heating \citep{Johansson+09}. The theoretical aspects of the heating mechanisms are described in Sections \ref{etg-p.3} and \ref{etg-p.7}.

\citet{Diehl-Statler_08} reported the first systematic study of 36 ETGs. More recently, \citet{Kim+20} derived temperature profiles for a sample of 60 ETGs using the data products of the \chandra Galaxy Atlas (CGA, \citealt{Kim+19a}).  They grouped them into six representative profile types (see Figure \ref{fig.07}). They are, in order of the number of galaxies included, `hybrid-bump' (rising at small radii and falling at large radii), `hybrid-dip' (falling at small radii and rising at large radii), `negative' (falling all the way), `positive' (rising all the way), `double-break' (falling at small radii, rising at intermediate radii, and falling again at large radii), and `irregular'. The most common type is hybrid-bump. Together with the double-break type, they comprise 50 per cent of the studied profiles. The main characteristic feature is that the temperature peaks at an intermediate radius, roughly a few per cent of $R_{\rm VIR}$ (the virial radius), and decreases both inward and outward from the peak. This behavior resembles that of the temperature profiles of galaxy groups and clusters, where however the peaks of $T$ are observed at relatively larger radii, $\sim$10\% of $R_{\rm VIR}$ \citep{Vikhlinin+05,Sun+09}.

Further considering the characteristic temperature and radius of the peak, dip, and break (when scaled by the gas temperature and virial radius of each galaxy), and the observational limitation at the outskirts of the galaxies, \citet{Kim+20} proposed a universal temperature profile that may explain 72 per cent (possibly up to 82 per cent) of the ETG sample (Figure \ref{fig.08}). In this scheme, $T_{\rm X}$ peaks at $R_{\rm MAX} = 35 \pm 20$ kpc (or $\sim$0.04 $R_{\rm VIR}$) and declines both inward and outward. The temperature dips (or breaks) at $R_{\rm MIN}$ (or $R_{\rm BREAK}$) $=3$--5 kpc (or $\sim$0.006 $R_{\rm VIR}$). The mean slope between $R_{\rm MIN}$ ($R_{\rm BREAK}$) and $R_{\rm MAX}$ is $\alpha = 0.3 \pm 0.1$ in $T_{\rm X} \propto r^\alpha$. The temperature gradient inside $R_{\rm MIN}$ ($R_{\rm BREAK}$) varies widely, indicating different degrees of additional heating at small radii. The hot core of some ETGs with hybrid-dip, double-break, or negative profiles may be related to recent star formation.

\begin{figure}[t!]
\centering
\includegraphics[width=0.95\textwidth]{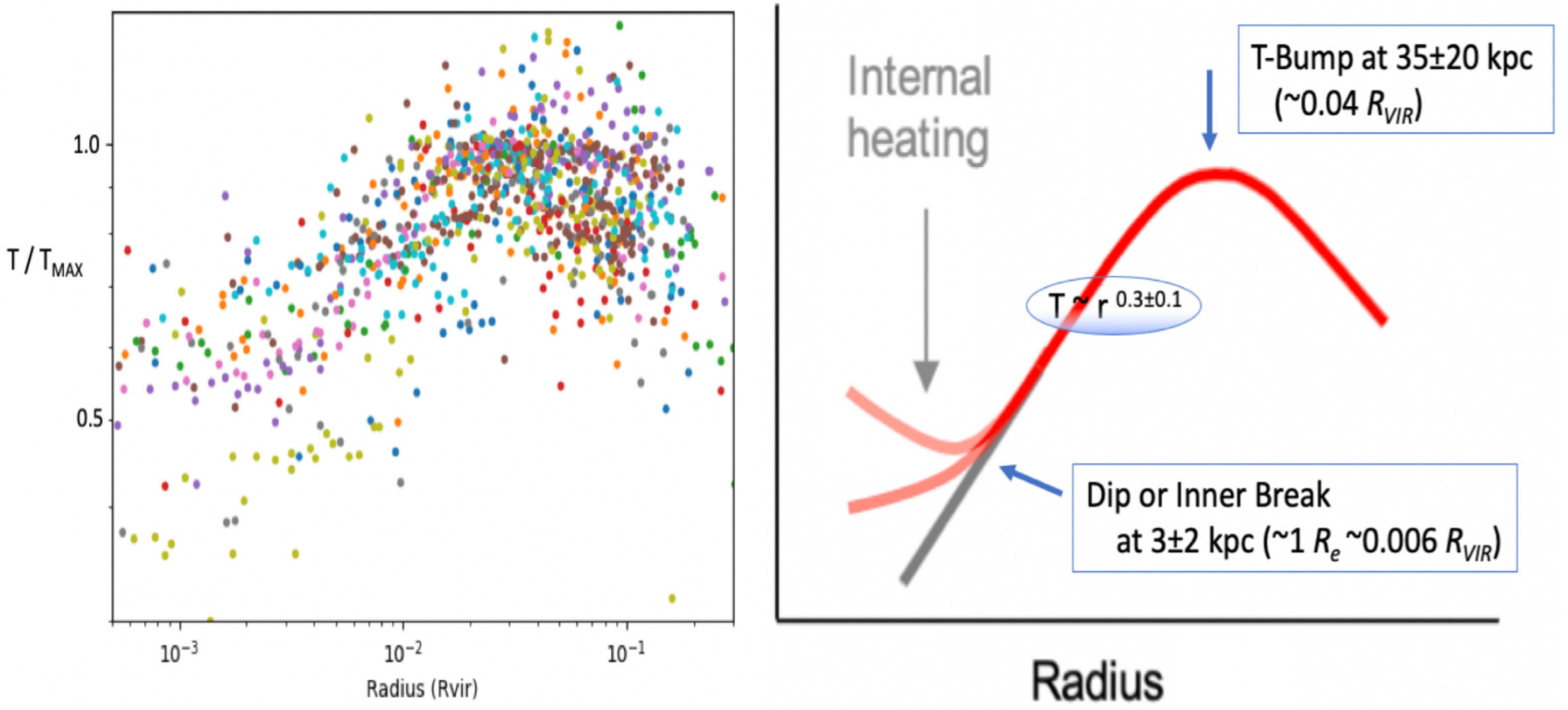}
\caption{Left: Temperature (scaled by $T_{\rm MAX}$) against radius (scaled by $R_{\rm VIR}$) for all galaxies in the hybrid-bump and double-break types. Right: A schematic diagram of the proposed `universal' temperature profile (taken from \citet{Kim+20}).}
\label{fig.08}
\end{figure}

\subsection{Radial Distributions of Fe Abundance}
\label{etg-k.5}
The Fe abundance profile is as important as the $T_{\rm X}$ profile, because it gives a window on the effect of stellar evolution on the hot ISM. However, Fe abundance is harder to measure and requires higher S/N data. Regarding various issues including multi-temperature gas, contamination by AGNs and LMXBs, background subtraction, resonance scattering, He sedimentation, and calibration uncertainty, we refer to the review in \citet{Kim-Pellegrini_12}.

Fe profiles were measured and reported primarily for hot-gas-rich systems. The typical profile shows that the Fe abundance is supersolar at the center of the galaxy and decreases with increasing radius, e.g., in NGC\,1399 \citep{Buote_02}, NGC\,5044 \citep{Buote+03}, NGC\,507 \citep{Kim-Fabbiano_04b}. We can call this declining Fe profile a `negative' type using the same terminology of the $T$ profile types in Figure \ref{fig.07}. For example, we show the Fe profile of NGC\,1550\ in Figure \ref{fig.09} (left panel). This negative type is expected because the Fe enrichment from SNe\,Ia  occurred mainly in the central region, and the Fe-enriched hot gas is slowly propagated to the outer region. High-resolution hydrodynamical simulations could qualitatively reproduce the negative gradient (see e.g., \citealt{Pellegrini+20}).

The second type of the Fe profile is one with a central Fe deficit. This Fe profile (rising at small radii and falling at large radii) is a `hybrid-bump' type. NGC\,4636 (Figure \ref{fig.09}, right panel) is an example (more can be found in \citealt{Rasmussen-Ponman_09,Panagoulia+15}). \citet{Panagoulia+15} showed that the central deficit is seen more often in galaxies with an X-ray cavity (as in NGC\,4636) and a shorter cooling time, and suggested that Fe may be incorporated in the central dusty filaments, which are dragged outwards by the bubbling feedback process. However, this measurement in the complex central region (typically $r < 10$ kpc) is challenging, and it may suffer from unknown systematic errors, e.g., resonance scattering (\citealt{Xu+02}; see also  \citealt{Gilfanov+87,Kim-Pellegrini_12}). The Fe deficit in the center needs to be confirmed by future missions (e.g., {\it XRISM}, expected to be launched in 2023).

\begin{figure}[t!]
\centering
\includegraphics[width=0.95\textwidth]{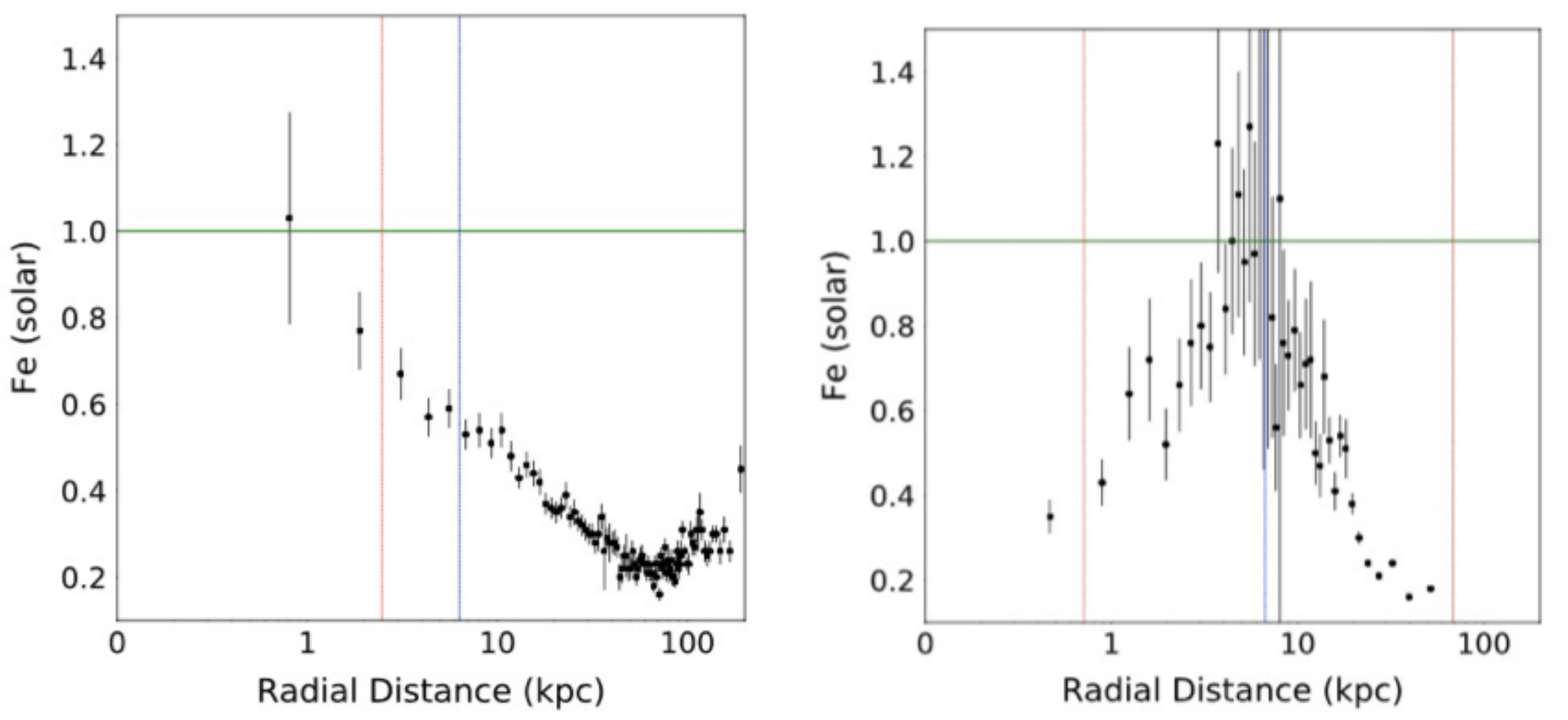}
\caption{Radial profiles of Fe abundance of the hot gas in NGC\,1550 (left) and NGC\,4636 (right). The blue vertical line denotes the half-light radius, and the dotted red vertical lines are at $r = 10$ arcsec and 16 arcmin, which indicate the inner and outer boundaries set by the \xmm point spread function and field of view. Note that the Fe abundance measurement at large radii ($r > 100$ kpc in NGC\,1550 and $r > 50$ kpc in NGC\,4636) is not reliable, even if the statistical error is small.}
\label{fig.09}
\end{figure}

\subsection{Entropy Profiles}
\label{etg-k.6}
Entropy ($K \simeq T n_e^{-2/3}$) has been used to address the thermodynamic history of the ICM. The entropy radial profiles of the ICM have been measured and analyzed (e.g., \citealt{Ponman+03}). The observed $K$ profiles were compared with theoretical predictions, e.g., $K \propto r^{1.1}$ in the absence of non-gravitational processes \citep{Voit+05}. The universal (or bimodal) $K$ profile and the presence (or absence) of the entropy floor in the ICM were debated in connection with the presence of H$\alpha$ emitting warm gas and cold molecular gas \citep{Cavagnolo+09,Panagoulia+14}. \citet{Babyk+18} investigated the entropy profiles of a sample of 40 ETGs. They show that ETGs have a higher rate of heating per gas particle compared to Brightest Cluster Galaxies, which may explain the lack of star formation in ETGs. We note that the hot ISM in ETGs is somewhat different from the ICM in terms of thermodynamics. The hot gas is primarily originated internally from stars, as opposed to the ICM with prevalent accretion or merging. Additionally, because the hot gas is rather complex in ETGs (as discussed in Section \ref{etg-p.3}), measuring density profiles in the hot ISM via deprojecting the surface brightness (with variable temperature, variable metal abundance, and clumpiness) is not straightforward, unlike in the relaxed ICM. 

\subsection{2D Spatial Distributions of X-ray Surface Brightness and Gas Temperature}
\label{etg-k.7}
The high spatial resolution \chandra images show a range of spatial features in the hot ISM, which was previously considered smooth and relaxed. The features seen in the X-ray surface brightness maps include X-ray jets (e.g., NGC\,315, \citealt{Worrall+03}), cavities coincident with radio jets/lobes (e.g., NGC\,4374, \citealt{Finoguenov+08}), nested cavities (e.g., NGC\,5813, \citealt{Randall+15}), cold fronts (e.g., NGC\,1404, \citealt{Su+17a}), filaments (e.g., NGC\,1399, \citealt{Su+17b}), and tails (e.g., NGC\,7619, \citealt{Kim+08}). These hot gas features give us insight on critical astrophysical processes, including the interaction of the AGN with the ISM of the host galaxy through radio jets/lobes, sloshing (see illustrative examples in Figs. 16--17 of \citealt{Markevitch-Vikhlinin_07}), ram pressure stripping of the hot halo interacting with the hotter ICM, and signatures of galaxy mergers. All of this is crucial for our understanding of galaxy formation and evolution via AGN feedback and environmental effects (e.g., \citealt{Kim-Pellegrini_12}, and references therein; see also Section \ref{etg-p.12}).

An excellent example of the importance of 2D maps is given by NGC\,3402 \citep{O'Sullivan+07}, where the X-ray surface brightness of the diffuse gas appears to be smooth and relaxed, but the $T_{\rm X}$ map clearly indicates a shell structure at 20--40 kpc, which is cooler than the surrounding gas (see Figure \ref{fig.10}). Another good example is NGC\,4649 (see the 2D $T_{\rm X}$ maps on the CGA website\footnote{\url{ https://cxc.cfa.harvard.edu/GalaxyAtlas/v1/cga_target_N4649_90601.html}}).  The surface brightness map of this galaxy shows that the hot gas distribution is smooth near the center but asymmetric in the outer region with extended tails (or wings) toward the northeast and the southwest. \citet{Wood+17} suggested that these wings may be caused by Kelvin--Helmholtz instabilities. The 2D $T_{\rm X}$ map further shows that while the cooler region is extended toward the same directions as seen in the surface brightness map, the asymmetrical features start from the center to the outskirts. This suggests a possible explanation by which the wings may be connected to the inner radio jets propagating toward the same directions, in addition to, or in place of, Kelvin--Helmholtz instabilities, which work at the outer surface.

\begin{figure}[t!]
\centering
\includegraphics[width=0.95\textwidth]{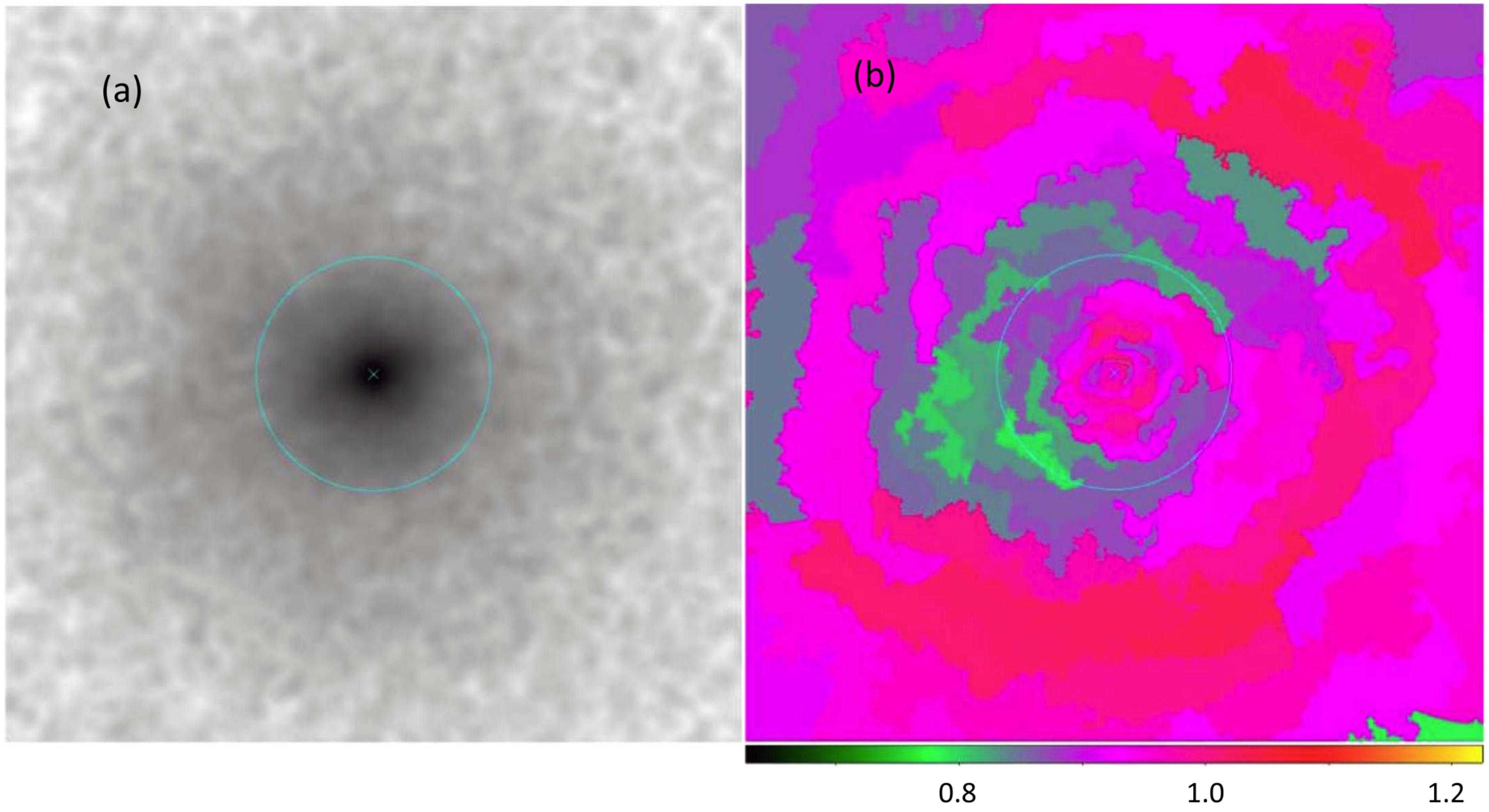}
\caption{(a) Surface brightness map (0.5--2 keV) and (b) $T_{\rm X}$ map of the hot gas in NGC\,3402. The cyan ellipse indicates the optical $D$25 ellipse (the 25 $B$-mag per square arcsecond isophote). Taken from \citet{Kim+19a}.}
\label{fig.10}
\end{figure}

The spatial distribution of the Fe abundance poses observational constraints on the metal enrichment history of the hot ISM. The phenomena at play include: the mass loss from evolved stars and SN explosions; the effect of internal mechanisms on the transport of the Fe, such as SN-driven winds and AGN-driven buoyant bubbles; and the effects of the interaction with the external environment, sloshing, and ram-pressure stripping. The \chandra and \xmm observations of NGC\,7619 provide a good example of Fe-enriched gas in the X-ray tail asymmetrically propagating to a large radius \citep{Kim+08,Randall+09}.

\citet{Islam+21} presented the Fe maps of 38 ETGs as a part of the \xmm Galaxy Atlas (NGA). In conjunction with the \chandra Galaxy Atlas, they compared a contrasting pair of ETGs, NGC\,4636 and NGC\,1550, to show how the Fe maps can be used to understand the metal transport processes. The \chandra and \xmm observations of NGC\,4636 revealed a complex hot gas morphology---cavities and a small-scale extension toward NW--SE at $r < 10$ kpc, an intermediate-scale ($\sim$20 kpc) extension toward WSW, and a large-scale ($\sim$50 kpc) extension toward N. The fact that extended gas directions are different at different radii is the typical phenomenon of sloshing as the center of the galaxy has been perturbed, or sloshed, more than once (see e.g., the simulations by \citealt{ZuHone+16}). The $T_{\rm X}$ and Fe abundance maps (see the spectral maps in the NGA website\footnote{\url{ https://cxc.cfa.harvard.edu/GalaxyAtlas/NGA/v1/nga\_target\_N4636\_90101.html}}) further show that the gas in the small-scale elongation, intermediate-scale extension, and large-scale enhancement is cooler and richer in Fe than the surrounding gas at similar distances. These spectral maps clearly show that the cooler, metal-enriched, low-entropy gas, originated from the stellar system by mass loss and SN ejecta, is stretched out primarily due to sloshing.

Unlike NGC\,4636, the hot gas in NGC\,1550 is relatively smooth, but the spectral maps exhibit exciting features that are not seen in the surface brightness map---the cooler and Fe-enriched gas is extended toward the E--W direction at $r < 15$ kpc (Figure \ref{fig.11}; see the Fe map in the NGA website\footnote{{ https://cxc.cfa.harvard.edu/GalaxyAtlas/NGA/v1/nga\_target\_N1550\_90301.html}}). Interestingly the E--W extension is aligned with the radio jet--lobe direction \citep{Kolokythas+20}, possibly indicating that the elongated cooler, metal-enriched, low-entropy gas is caused by the uplift by radio lobes, although sloshing may also occur in the different direction (see \citealt{Kolokythas+20,Islam+21}). The asymmetric Fe distribution (by the uplift), aligned with the radio jets and lobes, has previously been seen in a small number of clusters, the best example being Hydra A \citep{Kirkpatrick+09,McNamara+16}. NGC\,1550 may represent a nearby, smaller-scale example. For more examples of interesting spatially resolved features, we refer to the \chandra Galaxy Atlas \citep{Kim+19a} and \xmm Galaxy Atlas \citep{Islam+21}

\begin{figure}[t!]
\centering
\includegraphics[width=0.95\textwidth]{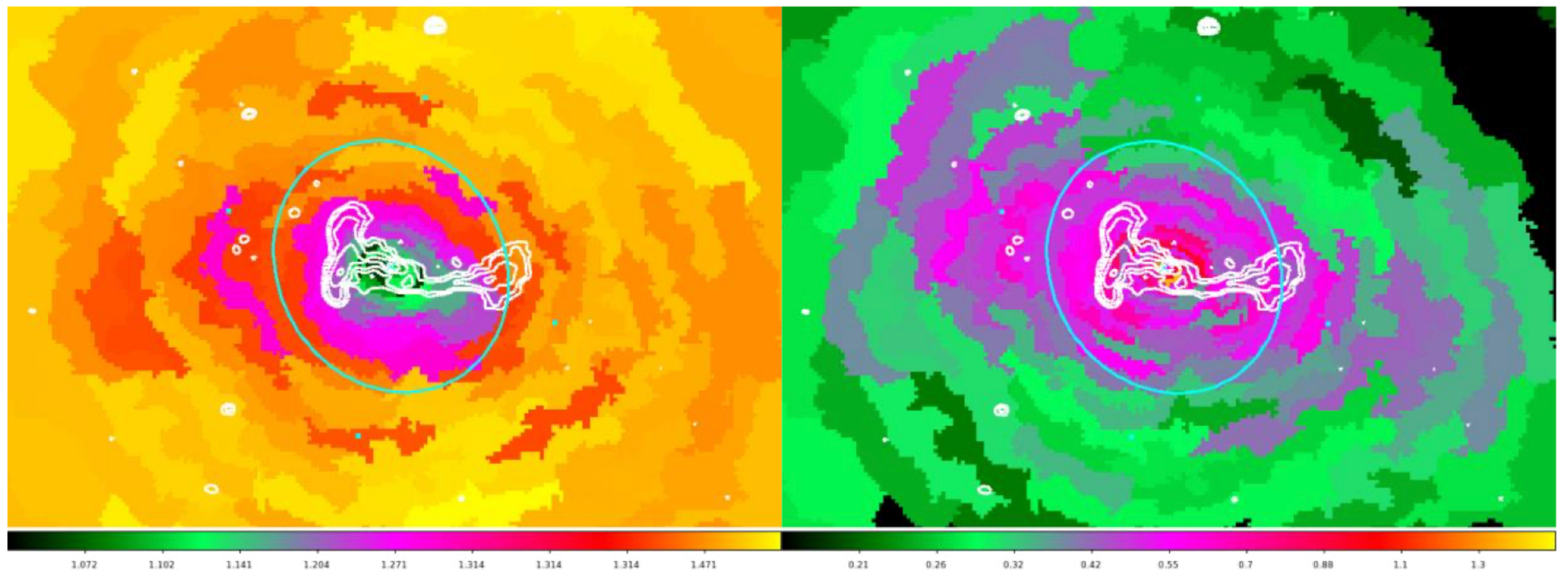}
\caption{{\it Giant Metrewave Radio Telescope} 610 MHz radio contours (from \citealt{Kolokythas+20}) showing the AGN jets and lobes, overlaid on the $T$ map (left) and Fe map (right) of NGC\,1550. The cyan ellipse indicates the $D$25 ellipse. Taken from \citet{Islam+21}.}
\label{fig.11}
\end{figure}

\section{Origin and Evolution of the Hot ISM in Early-Type Galaxies}
\label{etg-p}
We examine here the origin of the hot gas in ETGs, its heating sources, its dynamical status, its link with the galactic properties. In fact, the hot ISM has a tight relation with the host galaxy, being deeply influenced by its gravitational field, by various forms of mass exchange with it (as provided by stellar evolution, gas infall from the circumgalactic medium, and mass deposition via cooling), and by energy injections provided by SN explosions and accretion onto the central massive black hole. The following Sections deal in turn with these topics, and end with a global picture accounting for the observed $L_{\rm{X,GAS}}$ and $T_{\rm X}$ properties presented in Section \ref{etg-k}. We remind that in the following $L_{\rm{X,GAS}}$ is the galactic X-ray emission due to the hot gas only,  and $T_{\rm X}$ is the average temperature of the X-ray emitting gas.

\subsection{Origin of the Hot ISM }
\label{etg-p.1}
How do ETGs get their hot ISM? There are a few possibilities, dominating at different epochs. We take the view that ETGs are old systems, whose formation started at redshift $z>2$. At these early epochs, a major source for the galactic ISM was the infall of pristine gas within the DM halos; this gas was shock-heated to the virial temperature of the halos, that for halo masses of $M_{\rm{DM}}\gtrsim 10^{12}~\msun$ is $\gtrsim$ a few million degrees \citep{Mo+10}. Part of this gas cooled and contributed to star formation, a process that presumably stopped when the combined effects of AGNs and SNe heated the gas, and possibly also drove powerful outflows that displaced the ISM out to large radii, or even cleared the galaxy of it (e.g., \citealt{Naab-Ostriker_17}; Section \ref{sfg}). A hot circumgalactic corona (CGM) then formed, made of hot gas residing at large radii (out to the virial radius of the DM halo), further enriched by mass expelled from the galactic central regions; the CGM material could be falling back in at later times, making another source of galactic ISM (e.g., \citealt{Mo+10,Hafen+19}). At later epochs, the building of ETGs proceeded via minor merging, that affected mostly their external stellar envelopes, and was mainly dissipationless \citep{Oser+10}; major galaxy merging may also have occurred (see \citealt{Cox+06,Smith+18}; Sections \ref{sfg-4} and \ref{sfg-5} for its effects on the hot gas content). During this epoch, the normal aging of the stellar population also shed a significant amount of mass into the ISM. In summary, the main sources for the hot ISM of present-epoch ETGs include gas residual of the star formation phase; accretion from the CGM; and the internal gas production from stellar evolution.

\begin{SCfigure}[][t!]
\includegraphics[width=0.6\textwidth]{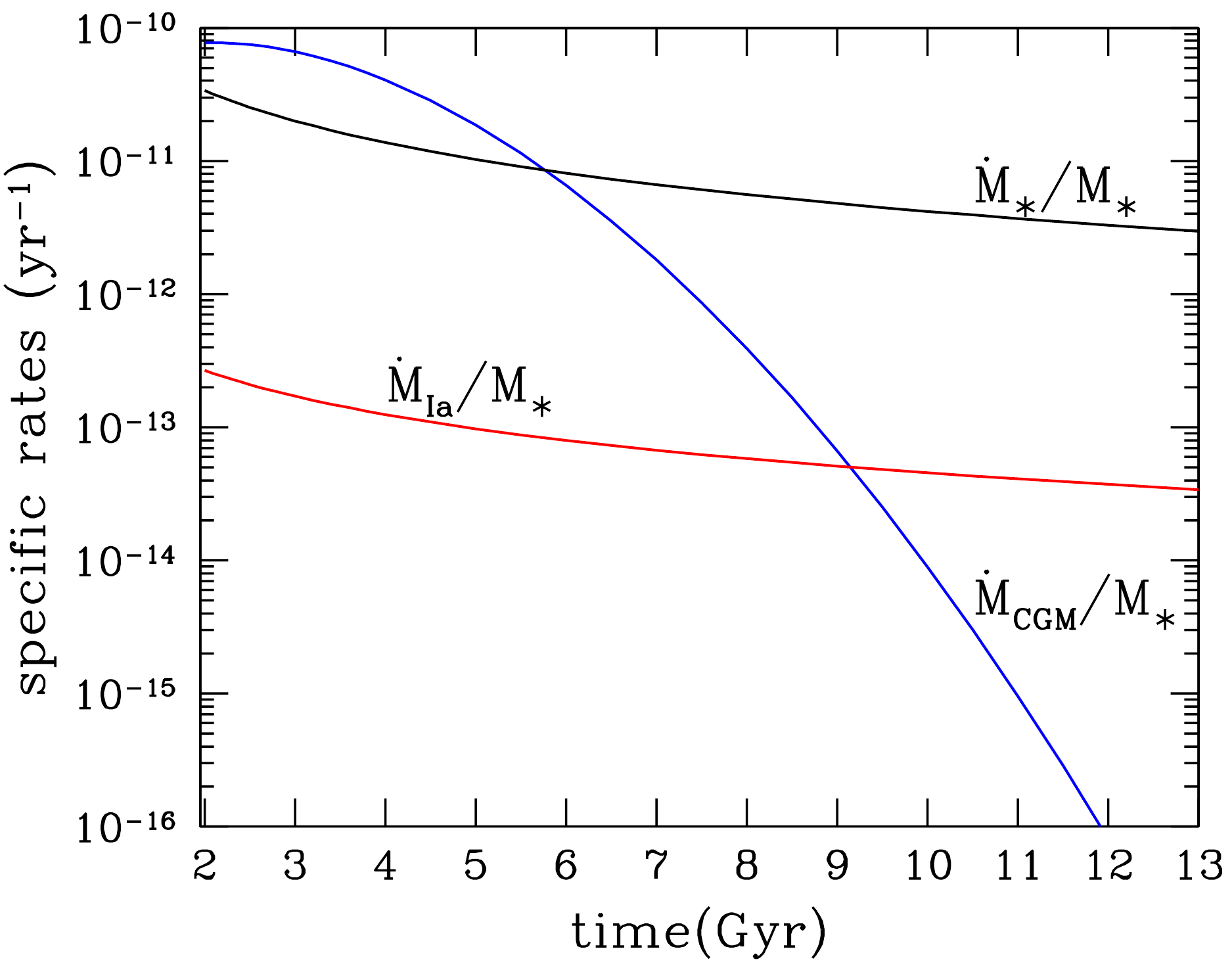}
\caption{The rates of mass input to the ISM described in Section \ref{etg-p.2}, normalized to $M_{\star}$. The aging stellar population inputs are $\dot M_{\star}$ (Equation \ref{eq.09}, in black), and $\dot M_{\rm{Ia}}$ (Equation \ref{eq.10}, in red). The CGM infall $\dot M_{\rm{CGM}}$ (in blue) is parameterized as suggested in Section \ref{etg-p.2}, with $t_0=3$ Gyr, $\Delta t$ equal to the present age of the Universe, and $M_{\rm{acc}}$ providing an integrated mass input from 3 to 13 Gyr of $0.1M_{\star}$.}
\label{fig.12}
\end{SCfigure}

\subsection{Relative Importance and Evolution of the Mass Sources}
\label{etg-p.2}
Estimates of the amount of hot gas remaining after star formation, and of the CGM material that falls in, are provided by cosmological simulations; they thus depend on assumptions about the complex physics of the galaxy formation and evolution, and on its implementation in the simulations. The residual gas mass should amount to a few per cent of the present-day stellar mass $M_{\star}$ (e.g., \citealt{Oser+10,Gan+19}), and subsequent evolution may well have caused its mixing and/or removal from the ETG, by the present epoch (Sections \ref{etg-p.10} and \ref{etg-p.12}). The rate of accretion from the CGM ($\dot M_{\rm{CGM}}$), though depending on poorly known factors, as for example the AGN feedback (Section \ref{etg-p.7}), was likely more important during the first few Gyr, and then declined sharply; the total accreted mass could be of the order of $\sim$10\% of $M_{\star}$ for massive ETGs. A suitable parameterization of this rate could be $\dot M_{\rm{CGM}}=2M_{\rm{acc}}\frac{e^{-(t/t_0)^2}}{1-e^{-(\Delta t/t_0)^2}} \frac{t}{t_0^2}$, where $M_{\rm{acc}}$ is the total accreted mass over the time interval $\Delta t$, and $t_0$ determines the time at which the rate peaks \citep{Gan+19}; an example of this parameterization is shown in Figure \ref{fig.12}.

The internal mass input provided by the stellar population includes stellar mass losses ($\dot M_{\star}$) and ejecta of Type Ia SNe (SNe\,Ia), $\dot M_{\rm Ia}$; type II SN events are confined to the first few Myr if the stellar population is overall old (Section \ref{sfg-3}). The most important contributions to $\dot M_{\star}$ are given during the red giant, asymptotic giant branch, and PN stellar evolutionary phases. In the hypothesis that the stellar population is `simple', i.e., born in a single episode, the collective return rate $\dot M_{\star}$ was estimated from stellar evolution theory; a robust approximation for $\dot M_{\star}$, after $\sim$2 Gyr of age, is:
\begin{equation}
\dot M_{\star} (t)=10^{-12}A \, M_{\star} (\msun) \, t_{12}^{-1.3}\quad \msun\,{\rm yr}^{-1},
\label{eq.09}
\end{equation}
where $A=3.3$ and 2.0 for the Kroupa and Salpeter IMFs, respectively, $M_{\star}$ refers to an age of 12 Gyr, and $t_{12}$ is the age in units of 12 Gyr (e.g., \citealt{Pellegrini_12}). The collective mass injection rate $\dot M_{\rm{Ia}}$ is much smaller, and can be calculated as the product of the mass contributed by one binary system exploding as SN\,Ia (1.4 $\msun$) times the event rate $R_{\rm SN}(t)$:
\begin{equation}
\dot M_{\rm Ia} (t)=1.4\msun\,R_{\rm SN}(t)=0.22\times 10^{-12} h^2 \, L_B(L_{B\sun}) \,  t_{12}^{-s}\quad \msun\,{\rm yr}^{-1},
\label{eq.10}
\end{equation}
where $R_{\rm SN}(t) = 0.16\, h^2 \times 10^{-12} L_B(L_{B\sun})\, t_{12}^{-s}$ (yr$^{-1}$) describes the evolution of the SN\,Ia explosion rate after $\sim$2 Gyr of age, $L_B(L_{B\sun})$ is the present-epoch $B$-band galactic luminosity, and $h$ is the Hubble constant in units of 70 km s$^{-1}$ Mpc$^{-1}$.  $R_{\rm SN}(t_{12}=1)$ gives the rate observed in the local Universe, and the slope $s$ parameterizes its past evolution with time; current estimates indicate $s\sim 1.0$ (e.g., \citealt{Maoz+14}). $\dot M_{\rm Ia}$ is then $\sim \dot M_{\star} /90$ at the present epoch, for a stellar mass-to-light ratio $M_{\star}/L_B=5.8$ appropriate for a Kroupa IMF at an age of 12 Gyr (see \citealt{Pellegrini_12} for more details).

The integration over  $\sim$10 Gyr of $\dot M(t)=\dot M_{\star}(t) + \dot M_{\rm Ia}(t)$ gives a total mass input of $\sim$0.05--0.1 $M_{\star}$, a figure close to that quoted above for the total accreted mass from the CGM; note, though, that the estimates for the stellar population are more certain, and, also important, that the two rates follow different trends with time (Figure \ref{fig.12}). Finally, the internal mass input rate follows the spatial distribution of the stellar population; thus, if the latter has a density distribution $\rho_{\star} ({\bf x})$, where ${\bf x}$ is the position vector, stars and SNe\,Ia inject mass locally at rates ${\dot{\rho}}_{\star} ({\bf x})\propto \rho_{\star} ({\bf x})$ and ${\dot{\rho}}_{\rm Ia} ({\bf x})\propto \rho_{\star} ({\bf x})$, respectively, and the hot ISM density distribution $\rho ({\bf x})$ has a local gas injection rate $\dot \rho  ({\bf x})={\dot{\rho}}_{\star} ({\bf x})+{\dot{\rho}}_{\rm Ia} ({\bf x})$. Note that $\rho ({\bf x})$ derived from observations, via the deprojection of the X-ray surface brightness (Section \ref{etg-k.4}), in general does not follow the spatial distribution of $\dot \rho  ({\bf x})$, because evolution of the hot gas and accretion from the CGM determine the total density profile (Sections \ref{etg-p.5}, \ref{etg-p.10}, and \ref{etg-p.12}).

\subsection{Heating of the Mass Sources}
\label{etg-p.3}
What heats the ISM to temperatures such that it radiates in the X-rays? Gas falling in from the outer regions of the galactic potential well is supposed to be shock-heated to the virial temperature of the dark matter halo, and reach temperatures of the order of $kT\sim 0.5$ keV in ETGs (Section \ref{etg-p.1}). Simple predictions for the properties of a hot ISM with this origin can be derived within the `self-similar' model, in which dark matter and baryons reach equilibrium under the action of gravity only (see Section \ref{etg-p.9}). Gas that is born within the galaxy, instead, as that originated by stellar mass losses, has very low temperatures when it is released by the stars. However, the interaction between the stellar ejecta and the surrounding ISM causes various shock and contact driven hydrodynamical and thermal instabilities, which result in the mixing of (most of) the ejecta with the ISM, and in its heating to approximately the temperature of the hot ambient medium, within few pc of the star, and on relatively short timescales \citep{Mathews_90,Parriott-Bregman_08}. In this process, the energy conveyed to the ISM is of the order of the thermalization of the random and ordered kinetic energies of the stars. Assuming a perfect mixing of the mass sources with the pre-existing gas, the energy input rate $L_{\sigma}$ for the random part of the stellar motions, over the whole galaxy volume $V$, is:
\begin{equation}
 L_{\sigma}\equiv \dfrac{1}{2}\int _V\dot\rho \, \mathrm{Tr} ({\boldsymbol{\sigma}}^2) dV, 
\label{eq.11}
\end{equation}
where ${\boldsymbol{\sigma}}$ is the velocity dispersion tensor of the stellar component. At the present epoch, $L_{\sigma}$ is of the order of $\sim$(1--5) $\times 10^{40}$ erg s$^{-1}$ in ETGs of central stellar velocity dispersion $\sigma_c=180$--260 km s$^{-1}$ (e.g., \citealt{Pellegrini_12}). The ordered stellar motions provide an energy input rate $L_v$ that depends on the difference in velocity between the streaming velocity of the stars (${\boldsymbol{v}}$) and the ISM velocity (${\boldsymbol{u}}$), and is calculated as:
\begin{equation}
L_{v} \equiv \dfrac{1}{2}\int_V  \dot\rho \,  {\left\lVert {\boldsymbol{v}} - {\boldsymbol{u}} \right\rVert}^2 dV. 
\label{eq.12}
\end{equation}
At variance with $L_{\sigma}$, this contribution cannot be estimated a priori, since it depends on the fluid velocity ${\boldsymbol{u}}$.  Of course, given the orbital nature of stars in ETGs, $L_{v} < L_{\sigma}$; furthermore, this inequality is likely to be large, because simulations indicate that ${\boldsymbol{u}}$ tends to be similar to ${\boldsymbol{v}}$ (see \citealt{Negri+14a} for more details).

SNe\,Ia contribute a major heating source for the ISM. The blast waves they originate dissipate their kinetic energy and form very hot bubbles that within a few million years disrupt by Rayleigh--Taylor instabilities, and share their thermal energy with the ISM \citep{Mathews_90}. Thus SNe\,Ia provide an energy injection rate, over the whole galaxy, of $L_{\rm SN} (t)=E_{\rm SN} R_{\rm SN}(t)$ erg yr$^{-1}$, where $E_{\rm SN}=10^{51}$ erg is the kinetic energy of one SN\,Ia event. At the present epoch, rewriting $L_{\rm SN}=5 h^2\times 10^{30} \, L_B(L_{B\sun})$ erg s$^{-1}$ (see Section \ref{etg-p.2}), one sees that $L_{\rm SN}$ largely exceeds $L_{\sigma}$\footnote{For example, for $\sigma_c=260$ km s$^{-1}$, and then $L_B=5\times 10^{10}L_{B\sun}$ from the Faber--Jackson relation, $L_{\rm SN}=2.5\times 10^{41}$ erg s$^{-1}$.}.
If a fraction of $E_{\rm SN}$ is radiated before the mixing of the ejecta is completed, a scaling factor $\xi \lesssim 1$ should multiply $L_{\rm SN}$ (Section \ref{sfg-3}). This factor is expected to be not much smaller than unity, because the SN\,Ia remnants evolve in a hot, low density medium \citep{Mathews_90,Tang-Wang_05}. Note that SNe\,Ia explode at random in the galaxy volume, with a frequency not high enough to have their hot bubbles overlap each other before dissolving into the ISM; this has implications for the spatial uniformity of the SN\,Ia heating, and the mixing of the iron content of each ejecta \citep{Mathews_90,Tang-Wang_05,Tang-Wang_10,Li+20}. Finally, the SN heating can also proceed via injection of cosmic rays or turbulence \citep{Pfrommer+17,Li+20}.

\subsection{Injection Temperatures and Observed Temperatures}
\label{etg-p.4}
The two heating sources $L_{\sigma}$ and $L_{\rm SN}$ are distributed over the galactic volume as $\rho_{\star}({\bf x})$, and locally provide the unit mass of injected gas an internal energy of $3kT_{\rm inj}({\bf x})/2\mu m_p$, where $T_{\rm inj}({\bf x})$ is the local mass-weighted injection temperature. 
$T_{\rm inj}$ is contributed by the thermalization of the motions of the gas-losing stars ($T_{\rm star}$), and of the velocity of the SN\,Ia ejecta ($T_{\rm SN}$), so that $ T_{\rm inj}=T_{\rm star}+T_{\rm SN}=(\dot M_{\star} T_{\star}+\dot M_{\rm Ia} T_{\rm ej} ) /\dot  M$ (e.g., \citealt{Pellegrini_12}). By approximating $\dot M_{\star}\approx \dot  M$, then $ T_{\rm star} \simeq T_{\star}$ and $T_{\rm SN}\simeq \dot M_{\rm Ia} T_{\rm ej} /\dot  M_{\star}$. Assuming that $E_{\rm SN}$ is entirely turned into heat,  $T_{\rm ej}= 2\mu m_p E_{\rm SN}/(3kM_{\rm SN})= 1.4\times 10^9$ K. Therefore $T_{\rm SN}$ is independent of the position in the galaxy, and evolves with time as $\dot M_{\rm Ia} /\dot M_{\star}$; at the present epoch, for a Kroupa IMF (Section \ref{etg-p.2}), $T_{\rm SN} \simeq 1.6\times 10^7$ K. $T_{\rm star}$ is instead independent of time with good approximation, but depends on the position in the galaxy, because $T_{\star} ({\bf x})$ is determined by the local stellar motions; for example, considering only the random part of the stellar motions, $T_{\star} ({\bf x})=\mu m_p \mathrm{Tr} ({\boldsymbol{\sigma}}^2 ({\bf x}))/3k$. The average mass-weighted $T_{\star}$ for the whole galaxy is:
\begin{equation} 
\langle T_{\star} \rangle =\frac{1}{3k}\frac{\mu m_p}{M_{\star}} \int_V \rho_{\star}\mathrm{Tr} ({\boldsymbol{\sigma}}^2) dV.
\label{eq.13} 
\end{equation} 
The integral in Equation \ref{eq.13} gives the kinetic energy $E_{\rm kin} $ associated with the stellar random motions, $E_{\rm kin}=0.5 \int dV \, \rho_{\star}\mathrm{Tr} ({\boldsymbol{\sigma}}^2) $, thus $\langle T_{\star} \rangle$ can be seen as the `stellar virial temperature'. An approximation for $\langle T_{\star} \rangle$ often used is $\langle T_{\star} \rangle \approx \mu m_p\, \sigma_c^2 /k$; this gives, for $\sigma_c=150$ and 250 km s$^{-1}$, respectively, $\langle T_{\star} \rangle\approx 1.7\times 10^6$ K and $4.7\times 10^6$ K (that is, 150 and 400 eV). In conclusion, at the present epoch,  $T_{\rm inj} ({\bf x}) = T_{\star}({\bf x})  + 1.6\times 10^7$ K, with the average  $\langle T_{\rm inj} \rangle= \langle T_{\star} \rangle + 1.6\times 10^7$ K,  where the term due to SNe\,Ia dominates. Observed emission-weighted temperatures for the hot ISM are smaller than $\langle T_{\rm inj} \rangle$, and closer to $\langle T_{\star} \rangle$ (Section \ref{etg-k.2}). Indeed, a significant part of the energy available to the gas that originates within ETGs, when it is injected, is conveyed in kinetic energy of the gas bulk flow, or in work done to uplift the gas in the galactic potential well (Section \ref{etg-p.10}), or in emitted radiation (Section \ref{etg-p.5}); it could also be that the energy transmitted by SNe\,Ia is lower than $L_{\rm SN}$ \citep{Pellegrini_11}.

The considerations above concern the {\it sources} of mass for the ISM, and illustrate how the various (cold) mass inputs experience heating processes expected to bring them to temperatures of a few million degrees. Note that, similarly to what pointed out for the gas density $\rho ({\bf x})$, also the gas temperature does not follow necessarily the $T_{\rm inj}({\bf x})$ distribution (as indeed observed; Section \ref{etg-k.4}). In fact, during its life, the ISM {\it as a whole} experiences additional sources of heating, as could be the compression resulting from the action of the gravitational field, thermal conduction from an external hot medium, and various forms of energy input from accretion on a central massive black hole (Section \ref{etg-p.7}). These processes are briefly illustrated below (Section \ref{etg-p.5}), after presenting the loss of energy via radiation, and the consequences it determines for the ISM evolution.

\subsection{Cooling and Evolution of the Hot ISM}
\label{etg-p.5}
Emission of radiation represents an energy loss for the hot gas that can be so important to influence its evolution and determine its properties. Since its temperature is of a few million degrees, and its density is typically low ($n_e\lesssim 0.1$ cm$^{-3}$), the hot ISM is highly ionized (with only the heavier species not fully ionized), optically thin, in a collisional ionization equilibrium state: the atoms are ionized or excited by collisions with electrons, and shortly after recombine or decay radiatively, emitting a photon that is lost by the gas \citep{Peterson-Fabian_06}. These radiative processes, plus thermal bremsstrahlung, produce an emissivity $\epsilon_{\nu}$ (the radiative power per unit frequency and per unit volume) that is written as $\epsilon_{\nu} =n_{\rm H}^2 \Lambda_{\nu}$, 
and $\Lambda_{\nu}(T,Z)$ is the cooling function that depends on the temperature and metal abundance $Z$. The definition of $\epsilon_{\nu}$ sometimes uses the electron number density $n_e$, or the total number density $n =\rho/(\mu m_p)$; the relations between the various number densities allow the conversion between definitions (for example, for the solar composition, $n=2.33 n_{\rm H}$, and $n_e=1.21 n_{\rm H}$). At the temperatures of interest for ETGs, $\Lambda_{\nu}$ has an important contribution from emission lines due to highly ionized neon, oxygen, magnesium, and silicon, and from the L-shell Fe transitions (see Figure \ref{fig.02}). The integration of $\Lambda_{\nu}$ over a frequency interval gives the emission in the corresponding energy band (e.g., the \chandra 0.3--8 keV sensitivity band); an example of the bolometric cooling function $\Lambda$ for various $Z$ is given in \citet{Sutherland-Dopita_93}. The hot gas luminosity in an X-ray band from a volume $V$ is then given by $L_{\rm{X,GAS}}=\int_V d^3({\bf x})\, n_{\rm H}^2({\bf x})\,\Lambda_{\rm X}(T({\bf x}),Z({\bf x}))$. 

The hot gas radiates and cools, especially in the central, more dense galactic region; the loss via radiation of thermal energy (which per unit volume is $3nkT/2$) takes place on a timescale known as the radiative cooling time $t_{\rm{cool}}=3nkT/(2 n_{\rm H}^2 \Lambda)$. Measured $t_{\rm{cool}}$ values can be as low as a few $\times 10^7$ yr or less, within a radius of $\sim$1 kpc \citep{Lakhchaura+18}. Across the region where $t_{\rm{cool}}$ is lower than the galaxy age (the central `cooling region'), if not heated to compensate for the radiative losses, the gas cools and is pushed in by the weight of the gas layers surrounding it; a cooling induced inflow is thus established. A simultaneous density increase within the cooling region is also produced, which makes $t_{\rm{cool}}$ shorter and shorter, until cooling becomes catastrophic in the innermost part of the galaxy; in fact, compressional heating, and even the SN\,Ia heating, cannot prevent rapid cooling (e.g., \citealt{Ciotti+91,Mathews-Brighenti_03}). This process is very similar to that described by the cooling flow model introduced for galaxy clusters \citep{Peterson-Fabian_06}, where though there are no distributed mass sources, and an infinite gas mass reservoir is present at the outer boundary of the cooling region. Cooling inflows are expected to be important in more massive ETGs (see Sections \ref{etg-p.10} and \ref{etg-p.11}).

\subsection{The Mass Deposition Problem}
\label{etg-p.6}
During unarrested cooling, the gas should cross the warm phase and end in the cold one; actually, the flow could be multiphase across all radii of the cooling  region (e.g., \citealt{Vedder+88}). The phenomenon by which mass is dropped out of the hot flow, and definitively becomes cold, is called mass deposition. The possibility for thermal instabilities to undergo nonlinear growth and condense the gas out of the hot flow \citep{Field_65} has been long debated, as they seem to be damped by conduction, ram pressure, or buoyancy, but instead may survive in presence of a magnetic field, and even be triggered by turbulent motions \citep{Loewenstein-Fabian_90,Pizzolato-Soker_05,Sharma+12,Gaspari+13}. In any case, if the gas radiatively cools down to low temperatures, the expected mass cooling rate predicted by the cooling flow model (${\dot{M}}_{\rm{cool}}\propto L_{\rm{X,GAS}}/T_{\rm X}$, \citealt{Peterson-Fabian_06}), or a modeling more specific for the hot ISM in ETGs (\citealt{Sarazin-White_88,Ciotti+91,Bregman+05}; Section \ref{etg-p.10}), is of the order of $\sim$1 $\msun$ yr$^{-1}$. Corresponding signs of mass deposition should be present, in the form of optical and UV filaments, molecular gas and dust, and signs of recent star formation; however, cooled or cooling material is observed at a lower (or far lower) level than expected. Exceptions are the central ETGs in groups and galaxy clusters, when surrounded by massive cooling flows \citep{Werner+14,McNamara+16}; some of them also suggest a link between the molecular gas content and the (shorter) cooling time \citep{Lakhchaura+18,Babyk+19}. In normal ETGs, optical emission lines with spatial coincidence with the X-ray cooling gas are seen in some cases (e.g., \citealt{Werner+14,Lakhchaura+18}), but X-ray, optical and UV spectroscopy in general indicates gas cooling rates lower than expected \citep{Bregman+05}. Also signs of past mass deposition are not consistent with accumulation that lasted for a few Gyrs: only some ETGs show H\,\textsc{i} and CO emission, and in a relatively small amount of 10$^6$--10$^8$ $\msun$ \citep{Werner+19}; recent or ongoing star formation is also insufficient \citep{Ford-Bregman_13,Kuntschner+10}. Therefore, empirical evidence requires the presence of a mechanism to compensate or prevent all the gas cooling expected in the simple modeling above; this could act via direct heating, and/or the displacement of the gas from the central regions.

To compensate for the cooling losses, the possibility of thermal conduction from the hotter, outer gas layers has been considered; however, it is unclear  whether conduction can be effective, since thermal conductivity could be much reduced, for example, by a tangled magnetic field that shortens the electrons mean free path. The existence of hot gas coronae embedded in a hotter ICM is taken as an indication that conductivity must be largely suppressed compared to the theoretical maximum (the Spitzer value) to avoid evaporation \citep{Sun+07,Sarazin_12}.

\subsection{AGN Heating}
\label{etg-p.7}
A promising way to avoid the growth of large cooled gas masses is the release of energy caused by accretion onto the central SMBH, a phenomenon known as `AGN feedback'; the accretion energy should be transferred to the ISM in an amount of the order of that lost in radiation \citep{Fabian_12}. This release of energy could be continuous, compensating isotropically for cooling, or intermittent and possibly also anisotropic. Intermittency matches well with AGN variability observed in various bands, and is linked to the idea of an activity cycle: gas cooling feeds accretion, the consequent energy injection heats and possibly also displaces the gas from the galactic center, so that cooling and accretion stop; since the gas inflow from the outer galactic regions, and the internal input of mass, continue, the density starts increasing again in the central part of the galaxy, cooling is resumed, and then accretion follows another time (e.g., \citealt{Pizzolato-Soker_05,Ciotti-Ostriker_07,Gaspari+12,Ciotti+17}).

The accretion output can take different forms, mainly including radiation, AGN winds \citep{King-Pounds_15}, jets \citep{Blandford+19}, and cosmic rays \citep{Guo-Oh_08}. Their importance is first of all measured by the respective rates of injection of energy ($L$) and momentum ($\dot p=\dot{M}v$) into the ISM, which are not completely known. Then, their effectiveness in stopping cooling, or in reheating, depends on how they interact with the ISM, which is also currently a subject of intense theoretical and observational study (e.g., Section \ref{etg-k.7}). A short summary is given below.

\subsection{The Various Forms and Effects of the SMBH Accretion Output}
\label{etg-p.8}
The radiative output consists of photons with total luminosity 
\begin{equation}
\Lbh = \epsem\Mdotbh c^2,
\label{eq.14}
\end{equation}
where $\Mdotbh$ is the mass accretion rate onto the SMBH; the radiative efficiency $\epsem$ ranges from a maximum of $\sim$0.1 for high accretion rates [$\dot{m}=\Mdotbh/{\dot{M}}_{\rm Edd} \gg 0.01$, where ${\dot{M}}_{\rm Edd} = L_{\rm Edd} /(0.1 c^2)$ is the Eddington rate], when a standard (cold, optically thick, geometrically thin) accretion disk forms, down to orders of magnitude lower values, when accretion is radiatively inefficient, hot, and a geometrically thick disk forms ($\dot{m} \ll 0.01$; \citealt{Yuan-Narayan_14}). The radiative output dominates the accretion output when $\dot{m} \gg 0.01$, in a regime known as quasar mode or radiative mode. Radiation transfers energy and momentum (the latter produced at rate $\dot{p}_{{\rm rad}}=\Lbh/c$) to the ISM, via Compton heating and photoionization heating, and through the gradient of the radiation pressure, which is contributed by dust absorption, photoionization opacity, and electron scattering (e.g., \citealt{Ciotti-Ostriker_07}).

The accretion output in the form of AGN winds and outflows ($\Lw$), and jets ($L_{\rm j}$), is known as the kinetic or mechanical mode, since it takes the form of mechanical energy \citep{McNamara-Nulsen_12,King-Pounds_15}. The rate of kinetic energy injection due to winds and jets can be written as: 
\begin{equation}
\Lw = \epsw \Mdotbh c^2, \quad\quad L_{\rm j}=\epsilon_{\rm j} \Mdotbh c^2,
\label{eq.15}
\end{equation}
where $\epsw$ and $\epsilon_{\rm j}$ are the efficiencies of mechanical energy generation with a wind and a jet, respectively (e.g., \citealt{Ciotti+09,Ciotti+17,Hardcastle-Croston_20}). Like $\epsem$, also $\epsw$ increases with $\dot{m}$, reaching a maximum presumably $\gtrsim 5\times 10^{-4}$ in the quasar mode, as indicated by observations and theoretical investigations (see \citealt{Harrison+18} for a review). Therefore $\Lbh$ is larger than $\Lw$ at high $\dot{m}$, and presumably also at low $\dot{m}$, where though $\epsem$ and $\epsw$ are less well known. It is acknowledged that, when present, jets ($L_{\rm j}$) are an important energy output, at all $\dot{m}$ \citep{Yuan-Narayan_14}; at $\dot{m} \ll 0.01$ they largely dominate over radiation ($L_{\rm j} \gg \Lbh$), and therefore this accretion regime is named `radio mode' (see also \citealt{Werner+19}). Winds are an important source of momentum (produced at rate $\dot{p}_{\rm w}=\dot{M}_{\rm w} v_{\rm w}$, where $\dot{M}_{\rm w} = 2\Lw/ v_{\rm w}^2$, with $v_{\rm w} $ the AGN wind velocity), which likely dominates over $\dot{p}_{{\rm rad}}$ at low $\dot{m}$, and is still significant at high $\dot{m}$. Moreover, thanks to their large opening angle, winds effectively impact the ISM, producing shocks that heat the gas, and push it outward;  during the high $\dot{m}$ accretion mode, these effects can reach distances of the order of $\sim$1 kpc \citep{Ciotti+17,Arav+20}. Jets interact with the ISM in a different way: they drill through the ISM, and discharge energy at larger distance from the SMBH, even outside the galaxies where radio lobes can be observed \citep{Vernaleo-Reynolds_06}. In ETGs residing in groups or clusters, jets often inflate bubbles of hot plasma within the ISM (Section \ref{etg-k.7}), and could couple to the ISM via a number of mechanisms, which include:  the work done by the expanding bubbles, which displace and uplift the ISM; the dissipation of shocks or sound waves that the jet/bubbles can drive into the ISM; the mixing of the lobe plasma, with the release in the ISM of streaming cosmic rays and magnetic energy; and the dissipation of turbulent motions originating from the flow around the rising buoyant bubbles (see the reviews in \citealt{McNamara-Nulsen_12,Werner+19}). Turbulence has also  been suggested to induce overdensities that develop into cold gas clumps, which in turn feed accretion onto the SMBH \citep{Pizzolato-Soker_05,Gaspari+13,McNamara+16}. Turbulent motions cause the broadening of the X-ray spectral lines, which could be observed by future high-resolution X-ray spectrometers.

\subsection{Modeling of the Hot ISM: the Simplest Model}
\label{etg-p.9}
In a simple hierarchical model of structure formation, where increasingly larger systems are formed under the action of gravity only, all structures from galaxies to galaxy groups to galaxy clusters are expected to be scaled versions of each other. A natural question is then whether  also the hot gas permeating ETGs, groups, and clusters shows a continuity of properties, which is embodied in scaling laws followed by all these structures. Actually, predictions exist for such laws, within a simple model where only gravity acts, and then the mass $M$ is the only parameter determining all the properties of a structure on a given scale (the self-similar model; \citealt{Kaiser_86}). If the gas evolves neglecting cooling, and other forms of heating other than those linked to gravitationally induced motions, it ends in hydrostatic equilibrium at an average temperature ($T_{\rm X}$) close to the virial temperature of the potential well where it has fallen. Its $L_{\rm{X,GAS}}$ and $T_{\rm X}$ scale then with $M$ as power laws: $T_{\rm X}\propto M^{2/3}$, $L_{\rm{X,GAS}}\propto T_{\rm X}^2$ (for bremsstrahlung emission), and then $L_{\rm{X,GAS}}\propto M^{4/3}$ \citep{Kaiser_86,Mo+10}. These expectations are not fully met by the observed gas in clusters, and are markedly broken by the hot ISM of ETGs, which shows instead much steeper $L_{\rm{X,GAS}}\propto T_{\rm X}^{4.5}$ and $L_{\rm{X,GAS}}\propto M^{3}$ relations (\citealt{Kim-Fabbiano_13}; Section \ref{etg-k.2}). Moreover, these relations are tight when restricting to ETGs that are massive and slow rotators, but become poorly defined, or disappear, for ETGs of medium-low luminosity $L_K$ \citep{Kim-Fabbiano_15}. The steeper $L_{\rm{X,GAS}}$--$T_{\rm X}$ relation indicates that gas is less X-ray luminous than predicted, at any given $T_{\rm X}$; an effect of this kind is produced if the gas density is lower than expected, which comes from a loss of gas, and/or a form of gas heating that prevented compression (e.g., \citealt{Voit_05}). Indeed, deviations from the simple self-similar model are easy to foresee in ETGs: the hot gas originated also within the galaxy, not only from cosmological infall, and underwent significant cooling and heating, as detailed in Sections \ref{etg-p.5}--\ref{etg-p.8}. The effects of these `non gravitational'  processes were naturally more dramatic in ETGs with a shallower potential well than groups and clusters. A more complex picture of the hot ISM evolution in ETGs that aims at reproducing its observed X-ray properties in the local universe is illustrated below. 

\begin{figure}[t!]
\centering
\includegraphics[width=\textwidth]{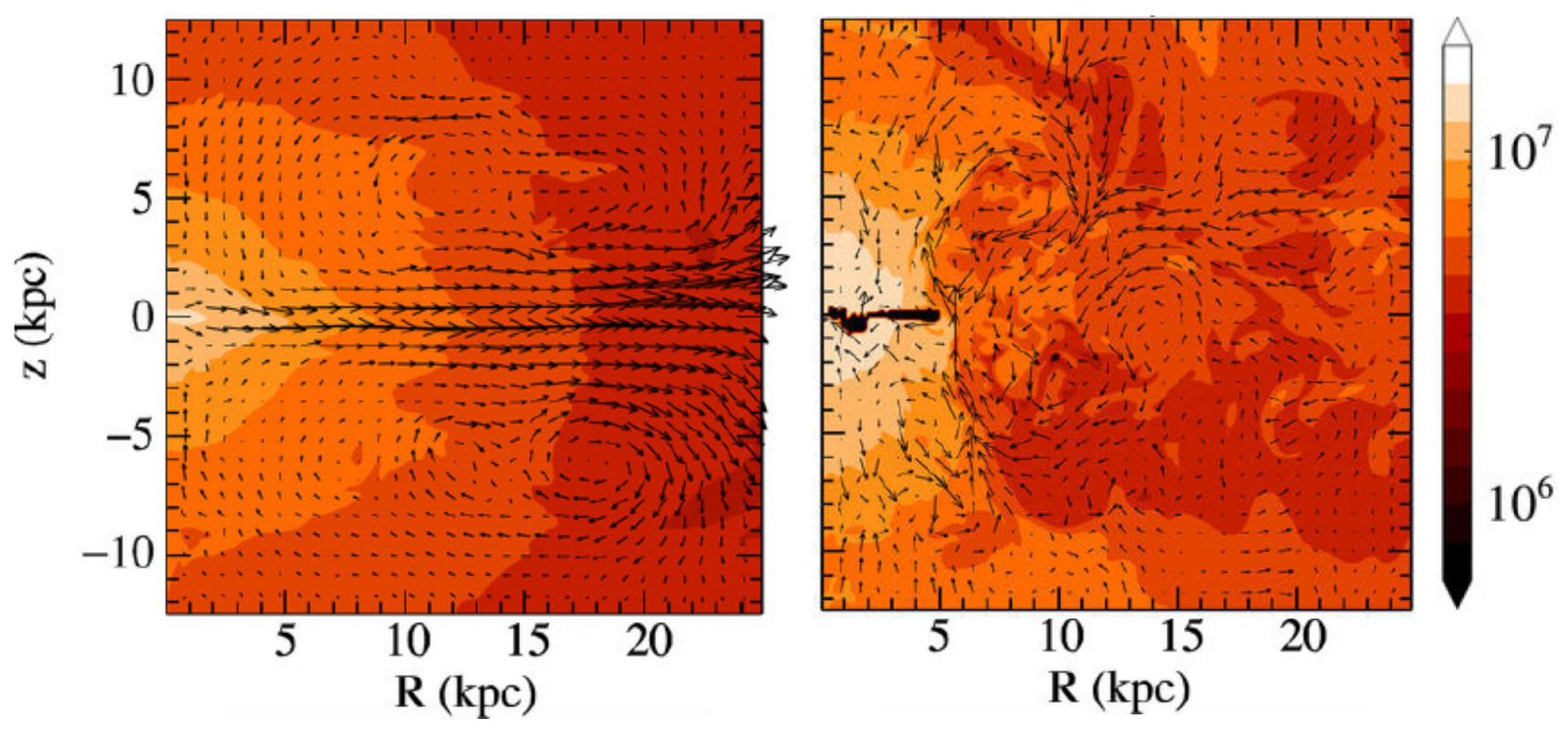}
\caption{Two examples of decoupled gas flows resulting from 2D hydrodynamical simulations, at the present epoch ($\sim$9 Gyr of age; \citealt{Negri+14a,Negri+14b}); colors indicate the temperature in K, on a meridional section, and superimposed arrows show the meridional velocity field, with the longest arrows corresponding to 170 km s$^{-1}$. The galaxy of $L_B=3\times 10^{10}L_{B\sun}$ has a flat shape, and is non-rotating (left), with a strong equatorial degassing, and rotating as an isotropic rotator (right), with the formation of a cold inner disk. Adapted from \citet{Negri+14a,Negri+14b}.}
\label{fig.13}
\end{figure}

\subsection{The Complex Lifetime of Hot Gas in ETGs}
\label{etg-p.10}
Before its discovery with X-ray observations, the absence of an observed ISM in ETGs was explained with the action of galactic winds. Assuming as mass source for the ISM only the ageing stellar population, the first models showed that the energy input from SNe\,Ia was large enough to drive the ISM in a steady outflow that escapes the galaxy (note that these models did not include dark matter halos; \citealt{Mathews-Baker_71}). The amount of hot gas in the galaxy is then very small, its $L_{\rm{X,GAS}}$ is very low ($L_{\rm{X,GAS}}< 10^{39}$ erg s$^{-1}$), and the total X-ray emission is dominated by stellar sources (Section \ref{etg-k.1}). Later, X-ray observations of ETGs revealed that their $L_{\rm{X,GAS}}$ spans a large range of values (Section \ref{etg-k.2}), from those expected for the collective emission of X-ray binaries (and then presumably corresponding to galaxies with the hot ISM in wind), up to $\sim$2--3 orders of magnitude larger values, thus incompatible with winds. The modeling then focused on the possibility of retaining the hot ISM within the galaxies, in a steady global inflow similar to the `cooling flow' described in Section \ref{etg-p.5} \citep{Sarazin-White_88}. This solution reproduced the largest observed $L_{\rm{X,GAS}}$ of non-central ETGs; however, in most ETGs $L_{\rm{X,GAS}}$ is lower than in global inflows. Intermediate $L_{\rm{X,GAS}}$ values could be explained with a gas flow that couples inflowing and outflowing regions (Figure \ref{fig.13}). The SN\,Ia heating keeps then the ISM mostly outflowing in galaxies with the lowest $L_{\rm{X,GAS}}/L_K$ ratio\footnote{This idea was confirmed by high resolution \chandra observations, which showed how the collective emission of X-ray binaries accounts for the total X-ray emission, in ETGs with the lowest $L_{\rm X}/L_K$ ratio (\citealt{Boroson+11}; Section \ref{etg-k.1}).}, and, for increasing $L_{\rm{X,GAS}}/L_K$, the inflowing region becomes more and more important. At the lowest $L_K$, all ETGs are mostly outflowing; at the largest, all ETGs host global inflows. These differences of the flow at the same $L_K$ are naturally produced by {\it observed} variations in the galaxy properties, even at the same $L_K$; these include, e.g., the shape of the stellar mass distribution (its flattening, Sersic index, effective radius), the central stellar velocity dispersion (indicative of the depth of the potential well), the importance of ordered rotation in the stellar kinematics, and the SN\,Ia rate \citep{Pellegrini_12,Negri+14a,Negri+14b}. For example, flatter ETGs are observed to have a lower $L_{\rm{X,GAS}}/L_K$ than rounder ones \citep{Negri+14b,Juranova+20}, which can be produced by a shallower potential well that favours the outflow. This is confirmed by hydrodynamical simulations, in low-mass ETGs; in medium- to high-mass ETGs, instead, it is a larger galactic rotation, which accompanies a flat shape, to produce a reduction of $L_{\rm{X,GAS}}$ (Figure \ref{fig.14}). In fact, the rotation of the stars is transferred to the stellar mass losses, and conservation of the angular momentum of the ISM injected at large radii causes the formation of a cold gas disk at lower radii; the hot gas density in the central galactic region is thus reduced. 
Indeed, slow rotators seem to have, on average, higher temperatures and higher $L_{\rm{X,GAS}}$ than fast rotators \citep{Sarzi+13}. Other causes of variation for $L_{\rm{X,GAS}}/L_K$ are environmental effects, as gas stripping and galaxy interactions that decrease the gas content \citep{Cox+06,Roediger+15,Smith+18}, and gas confinement due to an external medium, that increases $L_{\rm{X,GAS}}$, especially in central ETGs in groups and clusters (\citealt{Mathews-Brighenti_03,Sun+07,Sarazin_12}; see also Section \ref{etg-p.12}).

\begin{figure}[t!]
\centering
\includegraphics[width=\textwidth]{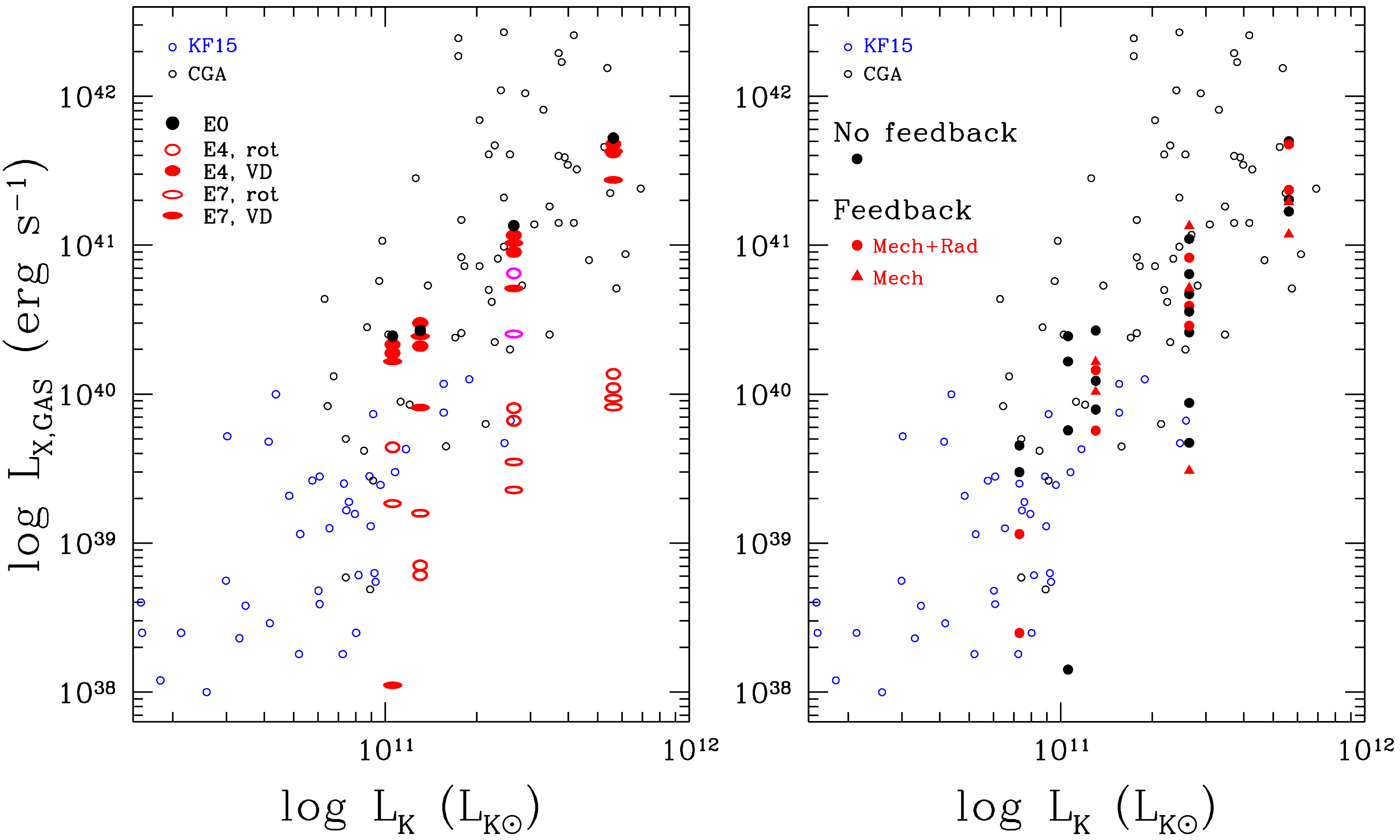}
\caption{Observed and model $L_{\rm{X,GAS}}$ in the 0.3--8 keV band, from within 5$R_{\rm e}$, versus the $K$-band luminosity $L_K$ of ETGs. Observed ETGs (open circles), with luminosity distance $D < 100$ Mpc, are from the \chandra Galaxy Atlas (CGA, in black; \citealt{Kim+19a}), and from \citealt{Kim-Fabbiano_15} (KF15, in blue). Model $L_{\rm{X,GAS}}$ values from hydrodynamical simulations refer to the present epoch. The left panel illustrates the possibility of varying $L_{\rm{X,GAS}}$ in galaxies that, at fixed $L_K$, are in all equal except for their shape and/or internal kinematics \citep{Negri+14b}: full black circles indicate spherical (E0) galaxies, filled and empty ellipses indicate respectively fully velocity dispersion supported (VD) galaxies, and isotropic rotators (rot); two magenta open ellipses show two cases of a lower rotational level \citep{Negri+15}. See Section \ref{etg-p.10} for more details. In the right panel the model galaxies, at fixed $L_K$, are in all equal but have been evolved without (full black circles) and with (red symbols) AGN feedback (mechanical, or mechanical and radiative); the models are described in \citet{Pellegrini+18}. See Section \ref{etg-p.12} for more details.}
\label{fig.14}
\end{figure}

\subsection{The Global Picture}
\label{etg-p.11}
The picture outlined above, supported by hydrodynamical simulations of the hot gas behavior, can account for the observed values of $L_{\rm{X,GAS}}$ and $T_{\rm X}$, and their trends with the galaxy properties, at least for relatively isolated ETGs (e.g., \citealt{Negri+14b}; Section \ref{etg-k.2}). In this view, the {\it average} $L_{\rm{X,GAS}}$--$L_K$ correlation (or the $L_{\rm{X,GAS}}$--$M_{\rm TOT}$ one, given that plausibly $M_{\rm TOT}$ increases with $L_K$) is fundamentally due to the increase of the binding energy per unit mass of the gas with  $L_K$, as implied by the observed Faber--Jackson relation, the steep scaling between luminosity and central stellar velocity dispersion $\sigma_c$ of ETGs \citep{Ciotti+91,Pellegrini_12}. However, $L_{\rm{X,GAS}}$ does not increase with $L_K$ in a tight and `self-similar' fashion, as if the flows were of the same kind at all $L_K$: at the lowest $L_K$, the flow is an outflow sustained by the SN\,Ia heating, while it becomes a global inflow at the largest $L_K$,  which produces the steep observed relations ($L_{\rm{X,GAS}}\sim L_K^3$, \citealt{Kim-Fabbiano_15}; and $L_{\rm{X,GAS}}\sim M_{\rm TOT}^3$, \citealt{Kim-Fabbiano_13,Forbes+17}). Superimposed onto the main effect of $L_K$, or
$M_{\rm TOT}$, there are secondary effects determined by other factors (the `shape', stellar kinematics, environment, etc.), which cause a range of $L_{\rm{X,GAS}}$ values at the same $L_K$. Importantly, these secondary factors have a stronger impact at medium--low $L_K$, where the gas is less bound, and they become less and less effective at the largest $L_K$, where global inflows are not easily `disturbed'; this explains the large scatter in the observed $L_{\rm{X,GAS}}$--$L_K$ (and $L_{\rm{X,GAS}}$--$T_{\rm X}$) relations at low and medium $L_K$, and their tightness at large $L_K$ (Section \ref{etg-k.2}; \citealt{Kim-Fabbiano_15}).

Being primarily dependent on the depth of the potential well, also  $T_{\rm X}$ shows correlations with $L_K$ and $\sigma_c$; however, temperatures are measured with larger uncertainties than luminosities, and therefore these correlations are poorer than those involving $L_{\rm{X,GAS}}$, and it is more difficult to derive the possible dependence of $T_{\rm X}$ on secondary factors. Note that measured $T_{\rm X}$ values are close to, or slightly larger than, the value of $\langle T_{\star} \rangle $, the `stellar' temperature (Sections \ref{etg-k.2} and \ref{etg-p.4}; \citealt{Pellegrini_11}); therefore, other forms of heating, as for example the AGN feedback, do not seem to affect the average $T_{\rm X}$. AGN feedback, indeed, is shown by numerical modeling to produce temperature fluctuations limited to the central region \citep{Pellegrini+12a,Gaspari+13,Li-Bryan_14}. More on the effects of AGN feedback as shown by numerical simulations is briefly summarized below, together with those of the environment.

\subsection{Two More Actors: Environment and AGN Feedback}
\label{etg-p.12}
For relatively isolated ETGs, the mass input from internal sources alone can account for $L_{\rm{X,GAS}} \lesssim (5$--$6)\times 10^{41}$ erg s$^{-1}$, while the largest observed values can reach $L_{\rm{X,GAS}} \sim 10^{42}$ erg s$^{-1}$ and more (Section \ref{etg-k.2}; Figure \ref{fig.14}). These, though, pertain to central or brightest galaxies in clusters and groups, for which it is difficult to separate the cluster/group emission from that of the ETG itself. Moreover, for galaxies at the center of a group, a cluster, or a substructure in it, the intragroup/intracluster medium has an important confinement effect, which significantly increases $L_{\rm{X,GAS}}$; gas accretion from outside can also be more important than in non-central ETGs \citep{Mathews-Brighenti_03}. Non-central ETGs, instead, may suffer from depletion of their hot gas content when moving in a dense external medium, due to the ram pressure stripping, or the Kelvin--Helmholtz instability caused by the velocity shears between the two media. Evidence of ongoing gas stripping is provided by upstream truncation of the hot coronae, and downstream gas tails, as observed in the Virgo and Fornax clusters (Section \ref{etg-k.7}; \citealt{Sarazin_12,Roediger+11,Roediger+15}). Infall of ETGs in the ICM is often accompanied by the phenomenon of `sloshing', which creates prominent cold fronts in this medium (Section \ref{etg-k.7}; \citealt{Roediger+11}). The environment has an effect also on the temperature profile in the outer galactic region, where the temperature tends to the large intragroup/intracluster medium values, when this medium is present, due to an external pressure effect \citep{Vedder+88}; this characteristic of the outer temperature profile is indeed observed (Section \ref{etg-k.4}).

Another important aspect concerning the hot gas evolution is that large cooling rates (and cooled masses) expected from inflowing gas must be prevented (Section \ref{etg-p.6}); and indeed, various forms of heating can originate from gas accretion onto the central SMBH (Section \ref{etg-p.8}), and are presumably at work (Section \ref{etg-k.7}). How much effective is this heating? Has it a role in the explanation of the observed X-ray properties ($L_{\rm{X,GAS}}$, $T_{\rm X}$), and does it have an impact on the scenario outlined in the previous Section \ref{etg-p.11}? Simulations of AGN feedback face the challenge of the large range of physical lengths involved, from the pc-scale at the center, to resolve the region close to the accretion radius of the SMBH \citep{Ciotti+17}, to the $\sim$100 kpc scale of the outer galactic regions. In terms of the high spatial resolution required, and the need to resort to 3D simulations, particularly challenging is the investigation of the effects of jets and bubbles. Therefore, in these cases, hydrodynamical simulations often zoom a limited spatial region close to the SMBH, and/or focus on a short time span, for example covering one activity cycle. The coupling efficiency with the ISM is in general assumed to be high, and jets are shown to provide an energy injection able to balance the cooling of the gas and maintain it in quasi-equilibrium (e.g., \citealt{Vernaleo-Reynolds_06,Gaspari+12,Li-Bryan_14,Yang-Reynolds_16}).

\begin{figure}[t!]
\centering
\includegraphics[width=\textwidth]{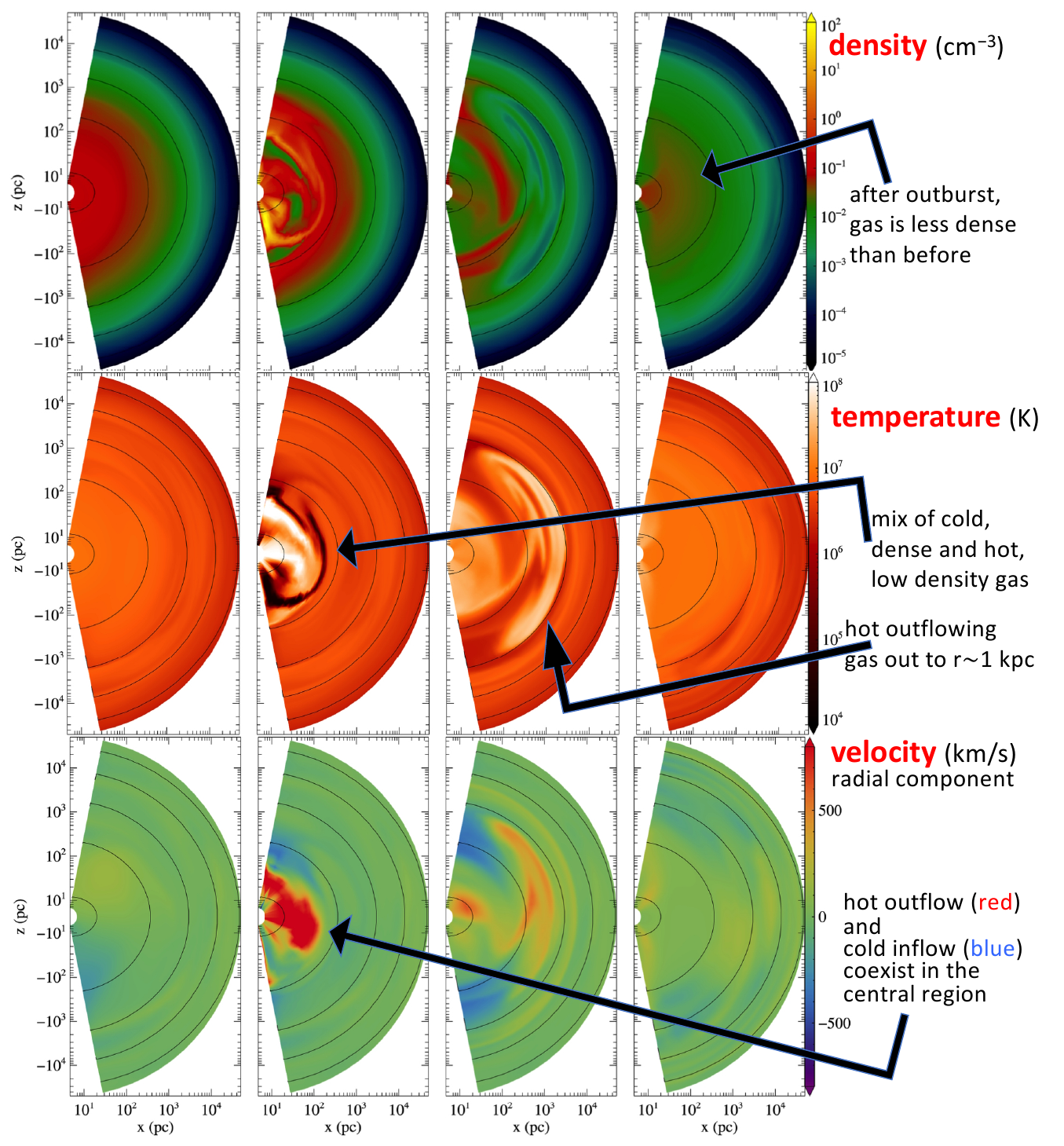}
\caption{Evolution of the gas density (upper panels), temperature (middle panels), and radial component of the velocity (lower panels) right before, during and after the end of a nuclear outburst caused by accretion, from the 2D hydrodynamical simulations of \citet{Ciotti+17}. From left to right, the four panels span a time interval of 0.1 Gyr. See Section \ref{etg-p.12} for more details. Adapted from \citet{Ciotti+17}.}
\label{fig.15}
\end{figure}

Investigations with pc-scale resolution close to the SMBH, and covering the last few Gyr of the lifetime, for a large set of representative model ETGs, were performed with 2D hydrodynamical simulations of the radiative and mechanical (due to AGN winds) feedback effects \citep{Ciotti+17}. It was found that the SMBH spends a very small fraction of time in a radiatively bright high-${\dot m}$ mode (with a duty cycle of $\sim$10$^{-2}$), while most of the time the accretion rate is small and the AGN is faint. The main, important effect of this AGN feedback is to reduce the amount of cooled gas, and prevent an excessive growth of the SMBH mass due to accretion. During outbursts, $L_{\rm{X,GAS}}$ is first increased and then drops down by a large factor within the central $R_{\rm e}$, because the gas is temporarily displaced from this region (Figure \ref{fig.15}); this behavior adds then another source of variation in $L_{\rm{X,GAS}}$ to those mentioned in Section \ref{etg-p.10}. The gas mass lost from the galaxy during its evolution is increased, but AGNs cannot clear (massive) ETGs from their hot ISM. As shown also by other simulations \citep{Pellegrini+12a,Gaspari+13,Li-Bryan_14}, the central temperature fluctuates as a consequence of nuclear outbursts (see also Figure \ref{fig.15}), which could explain the central cusps and dips frequently observed in the $T_{\rm X}$ profiles within few kpc from the center (Section \ref{etg-k.4}). Overall, when considering a large set of models, the effect of this type of AGN feedback on $L_{\rm{X,GAS}}$ and $T_{\rm X}$ seems marginal, and thus the range of the observed $L_{\rm{X,GAS}}$ is similarly reproduced, with and without this feedback (Figure \ref{fig.14}; \citealt{Pellegrini+18}). It is important to note, though, that these models do not include accretion from a CGM, and confinement from an external medium; if these are important, a larger feedback effect may be required to avoid excessive cooling.

\section{Future Prospects}
\label{fut}
With respect to the landscape of hot-ISM astrophysics, the forthcoming (or planned) space missions are undoubtedly expected to be very exciting. While \xmm and \chandra have significantly improved our knowledge of the properties of the hot ISM in both star-forming and early-type galaxies, in the medium to long term they are going be replaced by two other flagship observatories, which will extend their capabilities by one to two orders of magnitude, thus opening a new discovery space for X-ray astronomy: \athena \citep[e.g.,][]{Barret+20} and \lynx \citep[e.g.,][]{Gaskin+19}. The former was selected by the European Space Agency within its Cosmic Vision program, while the latter has been recently endorsed by the 2020 Astronomy and Astrophysics Decadal Survey of the National Academy of Sciences in the United States. These revolutionary facilities, however, will hardly become operative before the second half of the next decade. By contrast, the {\it eROSITA} telescope \citep{Predehl+21} aboard the \textit{SRG} orbital observatory \citep{Sunyaev+21}, launched in 2019, is performing an all-sky survey that will be 20 times more sensitive than the existing {\it ROSAT} maps once completed after four years. The new, all-sky data will be particularly useful to explore the outskirts of extended hot halos (in starbursts, mergers, and ETGs), and large-scale interactions with other galaxies and the underlying IGM (or ICM). 

{\it XRISM} \citep{Tashiro+20}, currently scheduled for launch in 2023, will provide extremely high-resolution ($\Delta E \leq 7$ eV) spectra thanks to the long-awaited, innovative technology of its X-ray microcalorimeter array (called {\it Resolve}). High spectral resolution is essential to tackle various open issues related to the abundance measurements of heavy elements and to investigate for the first time the dynamics of the hot, X-ray emitting gas \citep[e.g.,][]{Hitomi+17,Hitomi+18}. Although microcalorimeters are outperformed by gratings in terms of resolving power at low energies (i.e., long wavelengths), they ensure a huge improvement over CCD detectors. Separating the components of the K$\alpha$ triplet from He-like ions, for instance, can be a powerful method to probe the interaction between the hot starburst winds and the cold ISM, via charge exchange emission at the interface between the highly ionized and nearly neutral phases \citep{Liu+12}.

As we have pointed out earlier in this Chapter, spatial resolution is definitely the most critical characteristic of any X-ray instrumentation for the study of the hot ISM, not only for excising the point sources, but also to identify and inspect any substructures like clumps, filaments, and cavities. The sub-arcsecond resolution of \chandra will only be matched by \lynx (whereas \athena has a goal of $\sim$5 arcsec), which is expected to be also equipped with a spectrometer with resolving power $\lambda/\Delta\lambda \geq$ 5,000 and effective area $A_{\rm eff} \sim$ 4,000 cm$^2$ (as opposed to $\lambda/\Delta\lambda \leq 800$ and $A_{\rm eff} \sim 150$ cm$^2$ of the \xmm RGS). Smaller-class missions, based on either spatial or spectral resolution as those anticipated for \lynx, have also been proposed to NASA; if selected, they might fly within this decade, and considerably advance this research field. In any case, \athena and \lynx hold the greatest potential of transformational science. With their unprecedented sensitivity (afforded by the combination of larger collecting area and good angular resolution), these two observatories will finally reveal the emission from the most tenuous gas components of the hot ISM, such as the starburst superwind bubbles and the ETG halos well beyond 5 $R_{\rm e}$.

Simultaneously, future progress in numerical coding and an increased power of the computing resources will allow us to perform simulations with higher resolution, in terms of mass and length, with a larger dynamic range, covering a region extending from the nucleus to the circumgalactic environment, and elapsing for a simulated time that represents a significant part of the galaxy lifetime. A few important lines of investigation, not entirely new but addressed more adequately, could then be the following: 1) properly evaluate the mass accretion rate close to the SMBH (with $\sim$ pc scale resolution), and measure the effectiveness and the consequences of the various forms of AGN feedback, in the nuclear region and on the galactic scale; 2) include a more complete list of physical processes acting on the gas (as angular momentum, magneto-hydrodynamical phenomena, cosmic rays injection, turbulence); 3) follow the evolution of the various metal species, produced by star formation and SNe, and investigate possible inhomogeneities in their spatial distribution; 4) evolve the ISM in the galactic environment, by considering also the interaction with the circumgalactic medium; 5) map the hot ISM evolution for a set of model galaxies with different structural properties, representative of the whole galactic population.

\bibliographystyle{aa}
\bibliography{xISM}

\end{sloppy}

\end{refguide}

\end{document}